\documentclass[11pt]{article}%
\usepackage{color}
\usepackage{times,graphics,float}
\usepackage{enumerate}
\usepackage{amsfonts,amsthm,amscd}
\usepackage{amsmath}
\usepackage{amssymb}
\usepackage{float}
\usepackage{caption}
\usepackage[utf8]{inputenc}
\usepackage[english]{babel}
\usepackage[nottoc]{tocbibind}
\usepackage{eso-pic,calc,graphicx}
\usepackage{lipsum}

\theoremstyle{definition}

\usepackage{indentfirst}
\usepackage{multirow}
\usepackage{fancyhdr}
\usepackage{csquotes}
\usepackage[numbers]{natbib}
\usepackage{graphicx,subfigure}
\usepackage{setspace}

\author{Tingan Chen, Vic Patrangenaru}
\date{May 29, 2026}
\begin{document}
\begin{spacing}{1.0}

\begin{center}

\LARGE \bfseries  Extrinsic Analysis on BHV4\\[2cm]

\textsc{\large Tingan Chen$^1$, Garett Ordway$^2$, Vic Patrangenaru$^1$ }\\[1cm]

\textsc{\small Florida State University$^1$, Pensacola Christian College$^2$,
USA$^{1,2}$}\\[1cm]

\end{center}

\begin{abstract}
One investigates the extrinsic statistical analysis on the space of Billera-Holmes-Vogtmann tree space with four leaves ($\mathcal{T}_4$ or BHV4) based on its recently proposed novel representation (see \cite{ordway2025continuing})- the Spiky Projective Excavated Dodecahedron (SPED). Due to the symmetry of the SPED, the Veronese-Whitney (VW) embedding we consider here produces a natural extrinsic metric for a statistical analysis on BHV4. one derives the exact solution for the VW extrinsic mean and applies this novel method on a yeast genome dataset to study the phylogenetic trees of four distinct yeast clades.
\\[0.5cm]



\noindent E-mail : \\
tc19r@fsu.edu$^1$, vic@stat.fsu.edu$^1$, \\
gordway@faculty.pcci.edu$^2$

\end{abstract}
\end{spacing}

\section{Introduction}
Phylogenetic trees are trees whose structure represents a hypothesized evolutionary relationship between groups of organisms. Originally, these trees were built upon empirical evidence and scientific insights of biological scientists. Such objects are tremendously useful, if gathered either deterministically or randomly, benefiting the understanding of the evolutionary history in nature. In practice, a common procedure to build such trees computationally, is by clustering the DNA/RNA sequences based on a selected evolutionary distance.  While the resulting trees may not capture the order of appearance in the evolutionary history (the distances used for classification are symmetric by definition, and therefore can not take into account the direction of time), the combinatorial structure of these tree captures the similarity/dissimilarity of the DNA/RNA sequence of the leaf nodes, and will be useful to study hypothesis related to an animal/virus classification. Generically, the trees are restricted to be binary, i.e. each node has at most two ``children", therefore a such a tree with $n$ leaf nodes will have $n-2$ interior edges, which are edges that do not directly connect with a leaf node.

\begin{figure}[!h]
\centering
\includegraphics[scale=0.3]{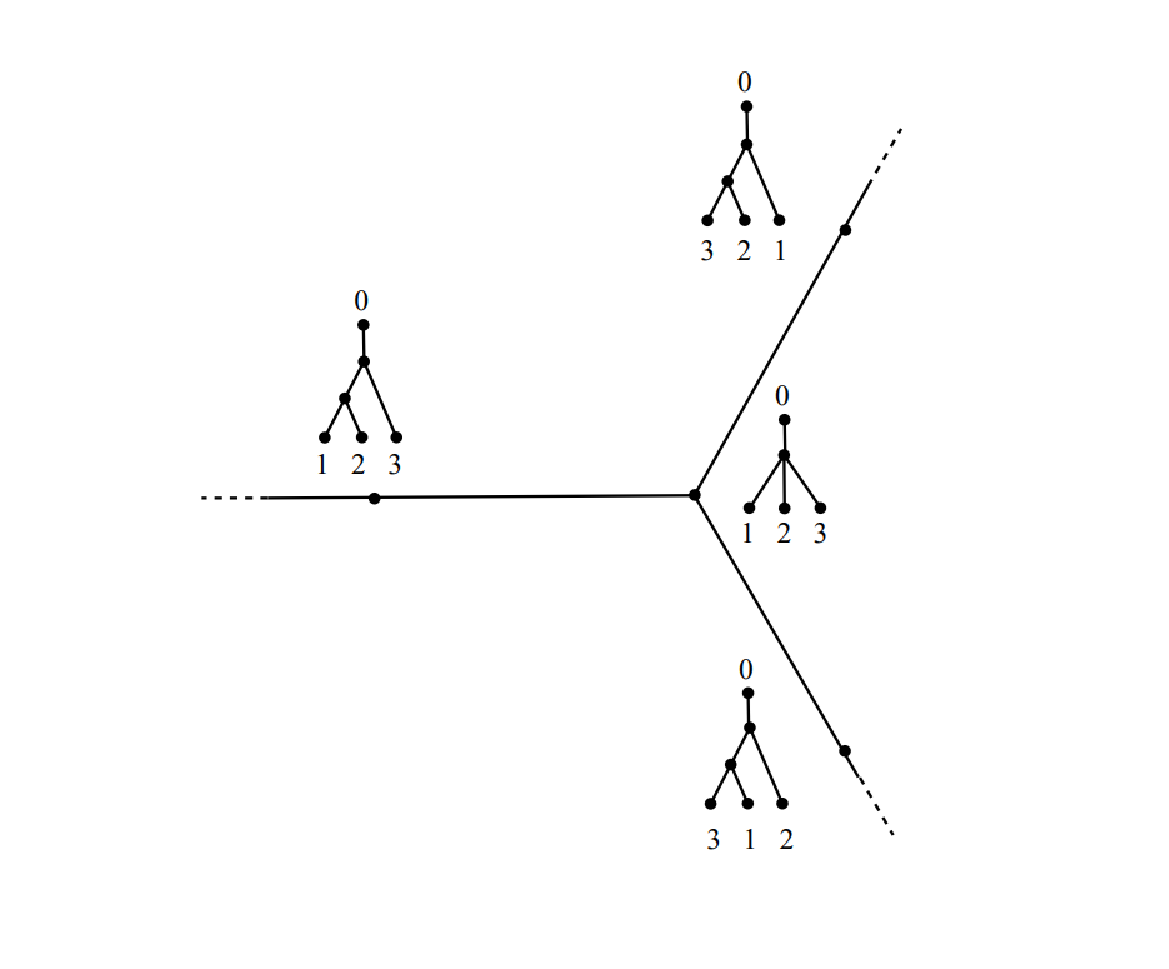}
\includegraphics[scale=0.3]{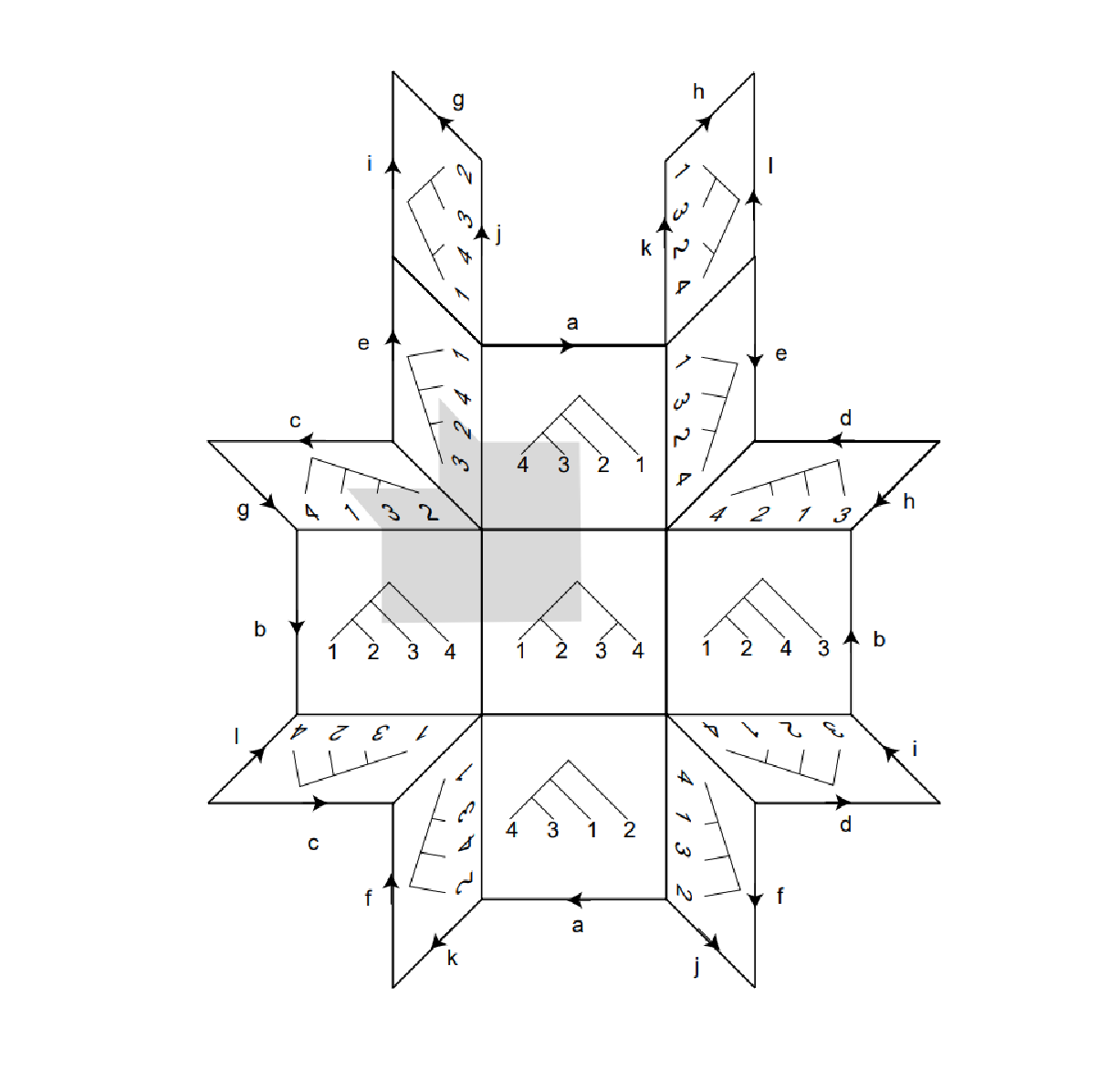}
\caption{ Representations of $\mathcal{T}_3$, and $\mathcal{T}_4$, respectively, from \protect\cite{BHV_tree}.} \label{fig:BHV_spaces}
\end{figure}

In \cite{BHV_tree}, a continuous metric space was defined for the set of all phylogenetic trees having a fixed set of leaves. In their work, the space of Phylogenetic trees with $N$ labeled leaf nodes, are named as $\mathcal{T}_N$.   figure \ref{fig:BHV_spaces} displays representations of rooted tree spaces $\mathcal{T}_3$, and $\mathcal{T}_4$.
Such spaces are naturally stratified spaces (\cite{Hotz_2013},\cite{patrangenaru2025nonparametric}). For $N\geq3$, the number of strata equals $(2N-3)!!$, which represents the number of distinct rooted binary trees with an unknown root, differentiated by their combinatorial structures (or sometimes referred to as the tree topologies). The length of the interior edges on each tree can also vary, which are represented by different points in the interior of each stratum. In the intersections of the strata, lie trees that can have node degrees higher than 2. For example, in the left picture of figure \ref{fig:BHV_spaces}, the origin of $\mathcal{T}_3$ represents the ``star tree", on which all of the 3 leaf nodes are from the same split, indicating a lack of significant DNA/RNA
mutations among the leaf sequences. Whereas on the interior of each interval (stratum), the trees have one interior edge.
Similarly, in the right picture of figure \ref{fig:BHV_spaces}, the corners of the squares represent the origin, i.e, the star tree of degree 4.

The space $\mathcal{T}_3$, is the simplest stratified space. It's elements are made up of three 1-dimensional (Euclidean) half-open, half-closed intervals, glued together at the closed end. On each stratum, the length of the one and only interior edge is represented as the distance to the origin. $\mathcal{T}_3$ leads to a 2D stratified space, the open book with three leaves. In general, one defines an open book by gluing  $d$-dimensional hyperplanes glued together by their common ``spine", obtained by identifying their $d-1$ dimension al boundaries (see for example, \cite{patrangenaru2025nonparametric}, Chapter 5).

\begin{figure}[!h]
\centering
\includegraphics[scale=0.4]{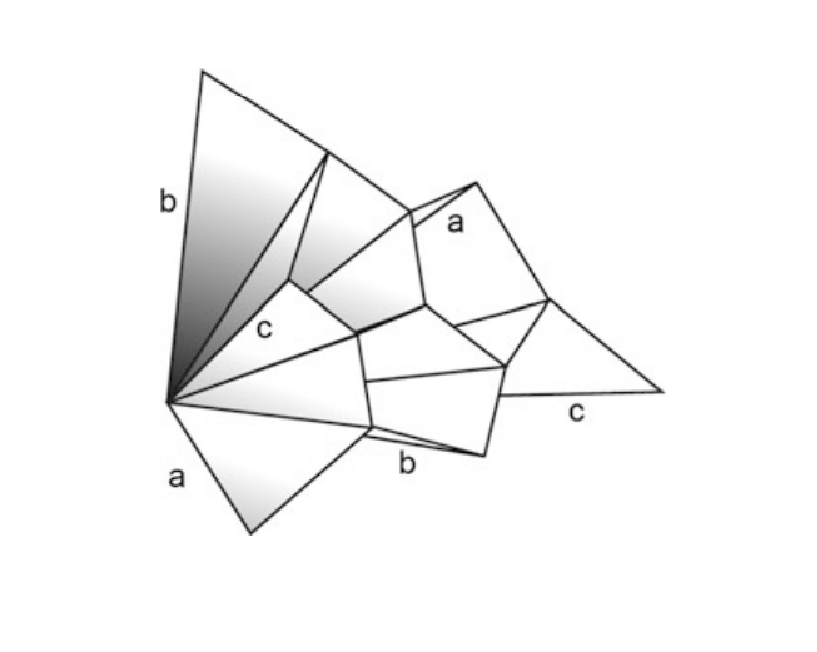}
\caption{3D description of $\mathcal{T}_4$ as a 2D stratified space. The edges with the same letters need to be identified, from (Billera et,al., 2001).}
\label{fig:BHV_w_cuts}
\end{figure}

\begin{figure}[!h]
\centering
\includegraphics[scale=0.4]{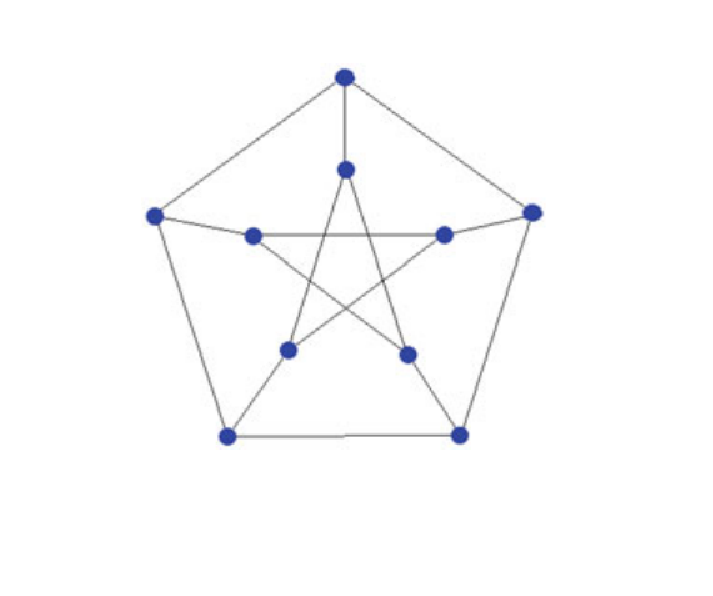}
\caption{Peterson Graph, a curve with no identification. However, this is only a one dimensional subset of $\mathcal{T}_4$ (which is an infinite cone of this object) and cannot be directly used for analysis.}
\label{fig: Peterson Graph}
\end{figure}

Compared with the space $\mathcal{T}_3$, the space $\mathcal{T}_4$ has a more intricate structure. As shown in the right picture of figure \ref{fig:BHV_spaces}, the space is made of 2D Euclidean quadrants glued together. Each tree type has four neighboring tree types, after considering the identification of the edges marked with the same letters. $\mathcal{T}_4$ can also be represented pictorially as in figure.\ref{fig:BHV_w_cuts}, which resembles a section of a beehive. Notice however, the edges with the same letters still need to identified. A more abstract representation of $\mathcal{T}_4$ is as cone over the Peterson Graph \cite{BHV_tree}, shown in figure \ref{fig: Peterson Graph}.


\begin{figure}[!h]
\centering
  \includegraphics[scale=0.4]{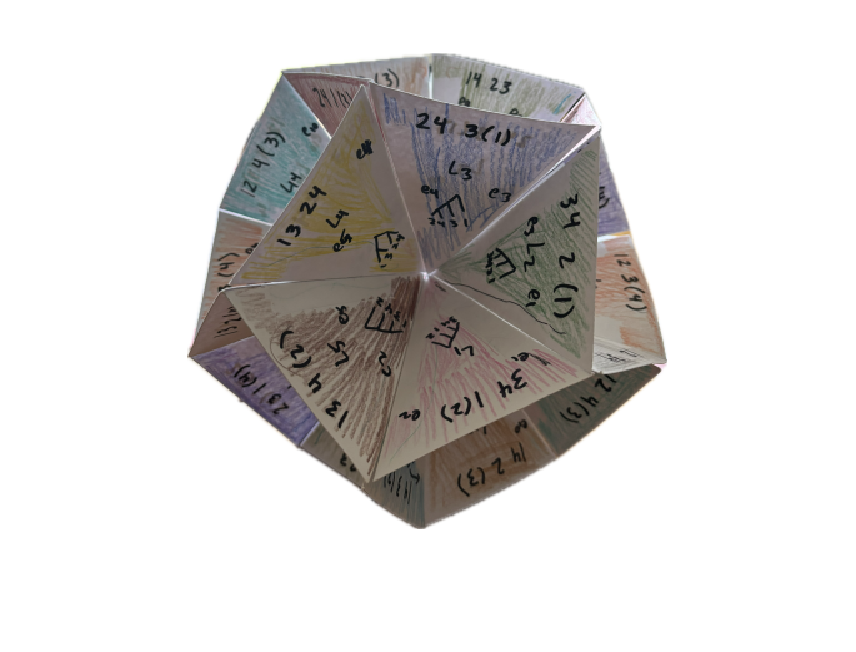}
  \hspace{1cm} 
  \includegraphics[scale=0.4]{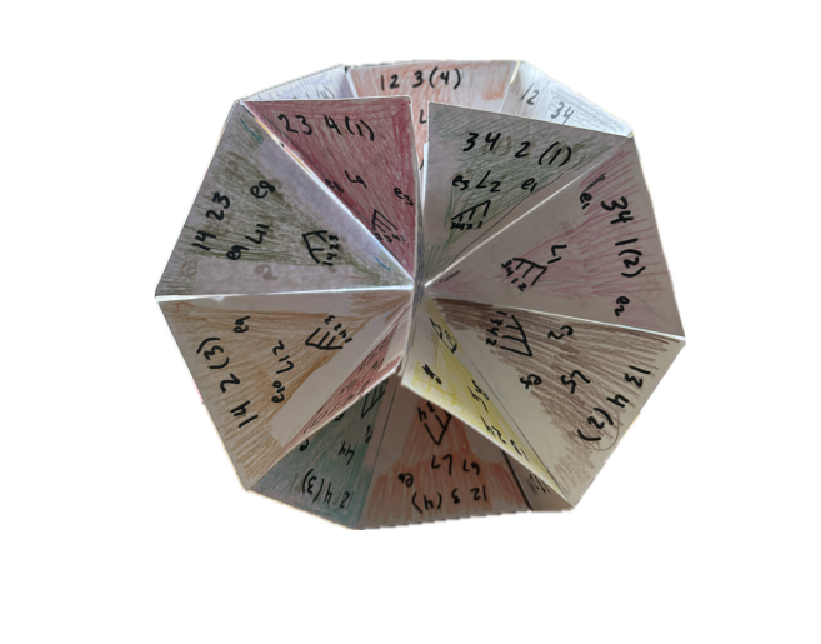}

  \caption{Views of half of $\mathcal{T}_4$ - PED \protect\cite{ordway2025analysis}: (Left) $\mathcal{Q}$5 view; (Right) open books view.}
  \label{fig:combined_views}
\end{figure}

To facilitate the analysis  on $\mathcal{T}_4$, \cite{ordway2025analysis} created a geometric representation, the Spiky Projective Excavated Dodecahedron (SPED). This model relaxed the right angles between each quadrant, by transforming the quadrant (square) into triangles. Therefore, all the quadrants can be glued together at the origin, which allows the analysis of data on all the quadrants simultaneously. See figure.\ref{fig:combined_views} for a display of this object. This model is named the Projective Excavated Dodecahedron (PED). The PED is obtained from an excavated dodecahedron, by identifying opposite points on it.

\begin{figure}[!h]
\centering
  \includegraphics[scale=0.15]{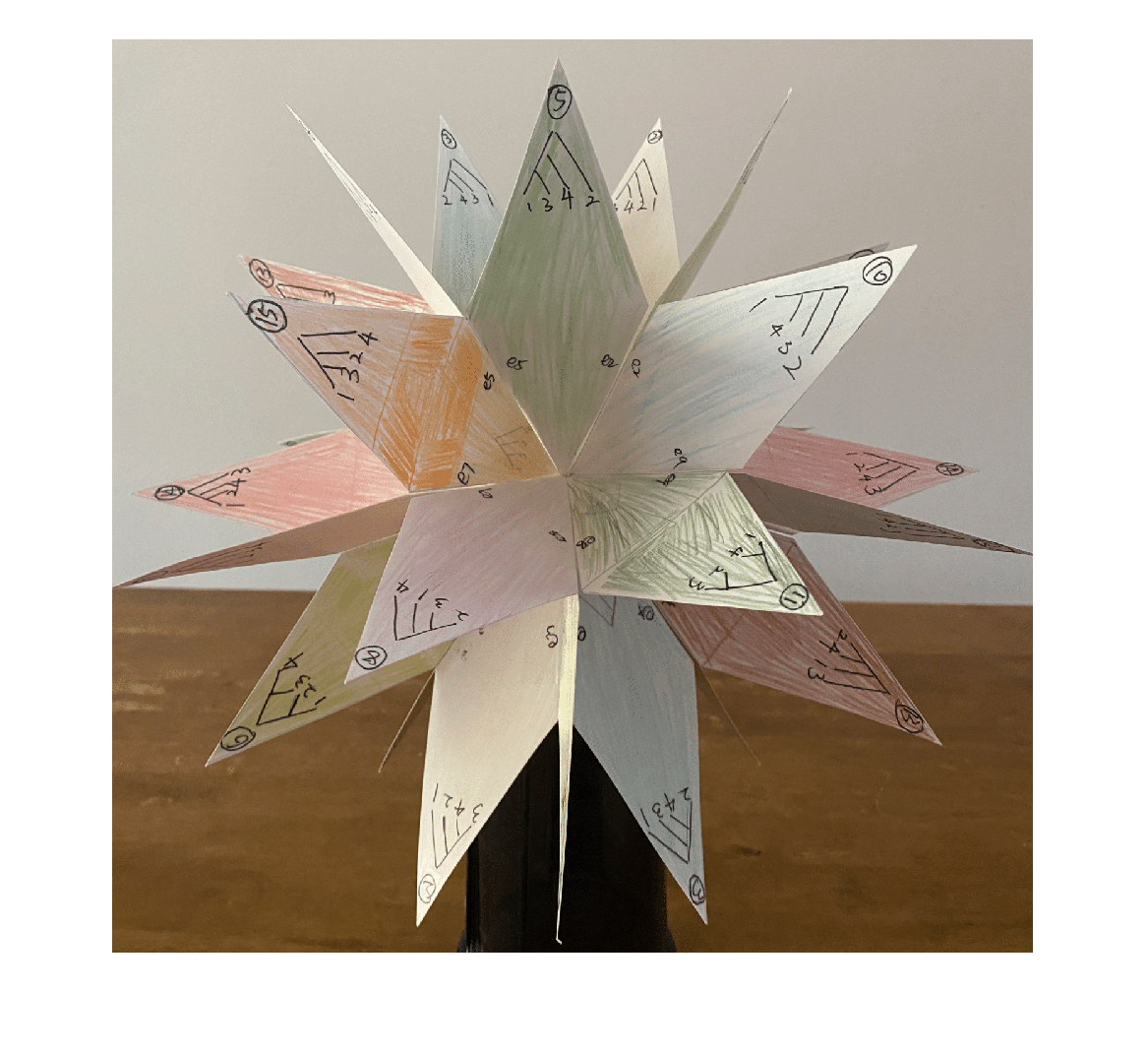}
  \hspace{1cm} 
  \includegraphics[scale=0.15]{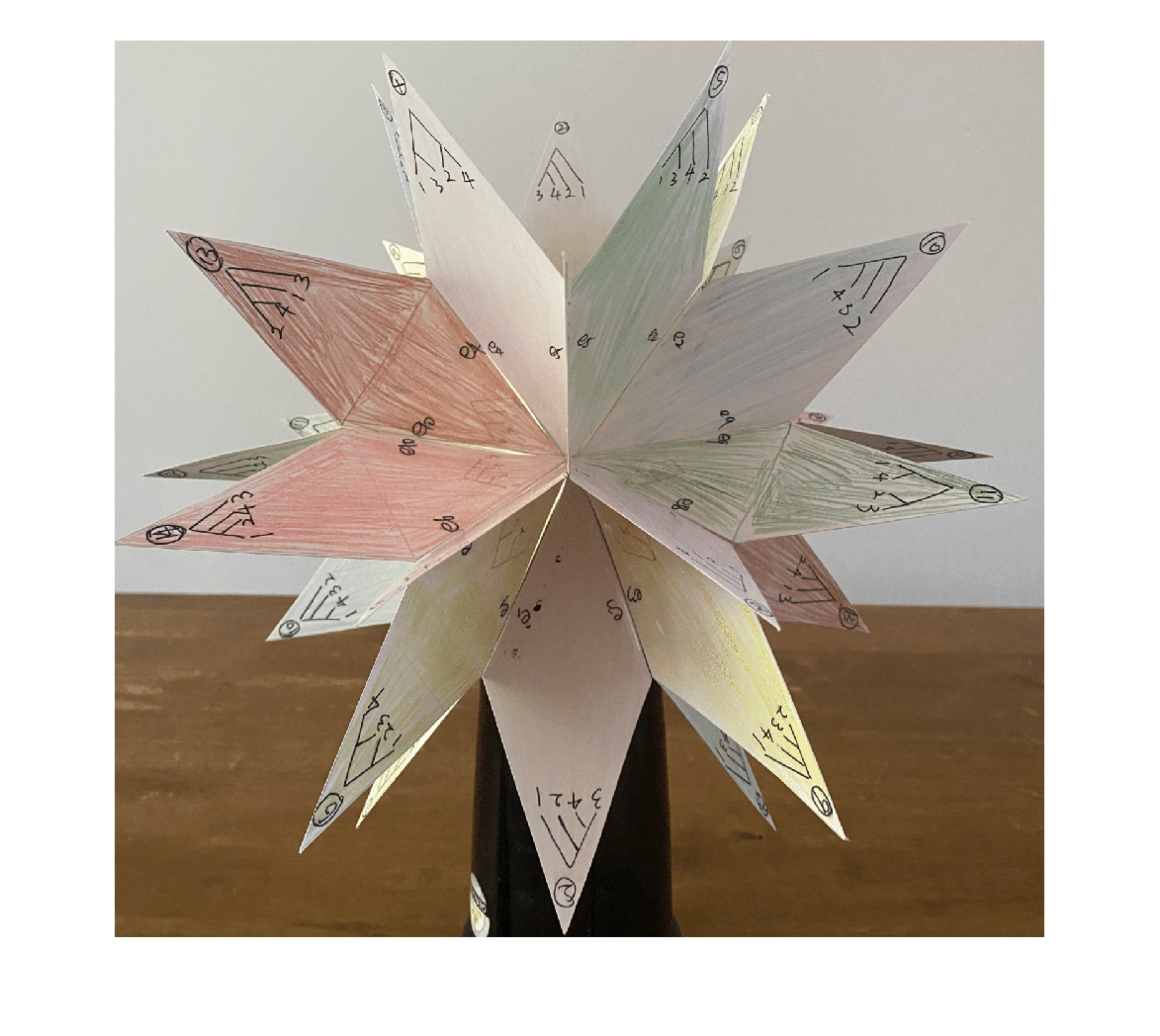}

  \caption{Views of $\mathcal{T}_4$ - SPED: (Left) $\mathcal{Q}$5 view; (Right) open books view.}
  \label{fig:combined_SPED_views}
\end{figure}

As seen from these figures, the PED is is made of triangles which represent halves of the original $\mathcal T_4$ quadrants, after a transformation. The edges of PED are identified correspondingly according to the rule in \cite{BHV_tree}. Therefore, PED is piecewise linearly homeomorphic to the original space of $\mathcal{T}_4$. However, it is more natural to deform the square blocks of $\mathcal{T}_4$ to rhombi, thus constructing a full Spiky Projective Excavated Dodecahedron (SPED), using piece-wise affine transformation mapping the squares into rhombi (diamonds). This model was first proposed in \cite{ordway2025analysis}, and applied later for exploratory analysis in \cite{ordway2025continuing}. A physical model of SPED is shown in figure.\ref{fig:combined_SPED_views}. Again the SPED, is obtained by identifying two opposite points on a spiky excavated dodecahedron; see figure \ref{fig:combined_SPED_views}.

\section{Method}

It is well known that the real projective plane $\mathbb RP^2$, also known as axial space, can be embedded into a numerical space for an extrinsic analysis. One of such embedding in $\mathbb R^4$ can be found in \cite{spivak1979comprehensive}; an equivariant embedding is preferred in axial data analysis; the most popular axial embedding is the Veronese Whitney embedding into the space of $3 \times 3$ symmetric matrices, $j:\mathbb RP^2 \to Sym_3$, where $j([x])=\frac{1}{x^Tx} xx^T, x \in \mathbb R^3\backslash 0.$ Axial data analysis using the resulting VW distance is extrinsic (see eg \cite{patrangenaru2016nonparametric}).

On the other hand, an intrinsic analysis using the arc distance between two axes, points on $\mathbb RP^2$ is often preferred as an intrinsic analysis, since the arc distance is the natural Riemannian distance on $\mathbb RP^2$. Computational challenges arising in an intrinsic analysis are met in \cite{jermyn2017elastic}. Theoretical advances in this direction can be found in, for example in \cite{dubey2019frechet} and \cite{hotz2024central}. However, the choice of such Riemannian metrics is also arbitrary, and the computational cost of such method can  be high \cite{bhattacharya2012extrinsic}, since the calculation of the intrinsic mean is often achieved via iterations, that may get stuck at a critical point of the Fr\'echet function, that is not necessary a point of minimum. The latter is an obstacle for downstream statistical inference is avoided if an extrinsic analysis is pursued.
In stratified spaces, there an additional downside of the intrinsic analysis, due to stickiness (see \cite{patrangenaru2025nonparametric}, \cite{Hotz_2013}). This phenomenon occurs in more diverse scenarios in $\mathcal{T}_4$, shown in \cite{Barden2013}, which essentially leads to a reduced sensitivity to deviations from the null hypothesis. {In general, extrinsic methods do not have this issue, see for example \cite{ellingson2014clt} and \cite{patrangenaru2025nonparametric}.}

We therefore embed the SPED flat surface into the space 3-by-3 symmetric positive semidefinite matrices using a Veronese-Whitney type embedding $j([x])=xx^T$, and derive the formula of the extrinsic mean extending the methodology for the extrinsic analysis on $\mathbb RP^2$ in \cite{mardia2005directions}.

\subsection{The Embedding of the SPED}
We parametrize a rhombus with vertices $x_1$ and $x_2$ as:
\[
\gamma(t_1,t_2)=t_1x_1+t_2x_2
\]
\begin{figure}[!h]
\centering
\includegraphics[scale=0.70]{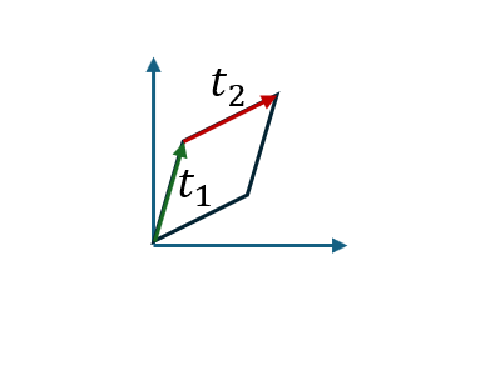}
\caption{Parametrization of rhombi in sPED}
\end{figure}

The embedded image of the parametrized triangle is: $\gamma(t1,t2)$ is $\gamma(t1,t2)\gamma(t1,t2)^T=$
\[
\eta(t_1,t_2) = t_1^2 x_1x_1^T + t_2^2 x_2x_2^T + t_1t_2 (x_1x_2^T+x_2x_1^T)
\]

\subsection{Displaying the Embedded Surfaces}
\begin{figure}[!htb]

\minipage{0.30\textwidth}
  \includegraphics[width=\linewidth]{SPED_parametrization.eps}
\endminipage\hfill
\minipage{0.16\textwidth}
\includegraphics[width=\linewidth]{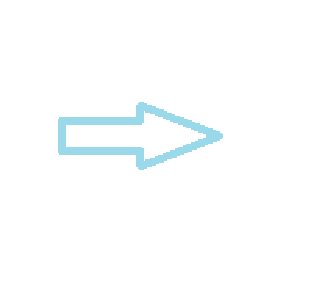}
\endminipage\hfill
\minipage{0.45\textwidth}%
  \includegraphics[width=\linewidth]{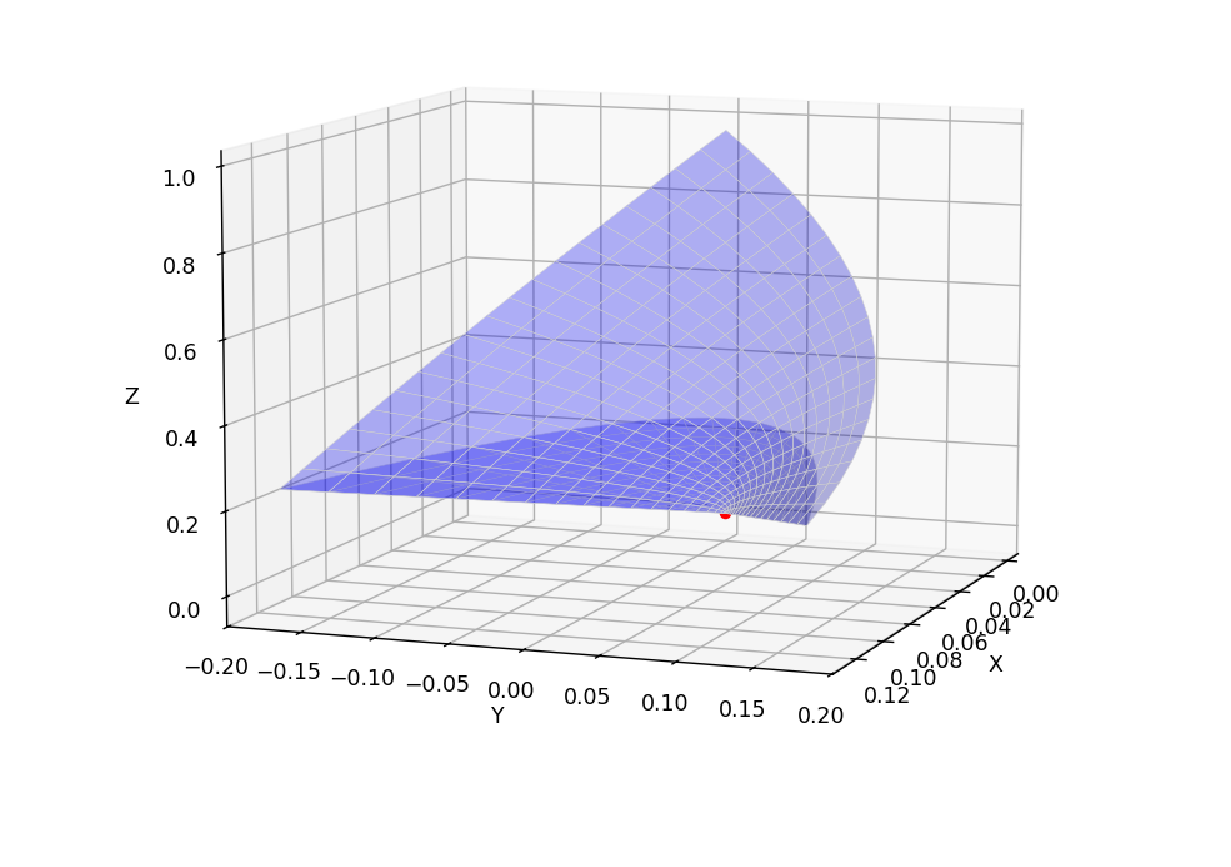}
\endminipage
\caption{Pictorial description the embedding of a surfaces. The resulting surfaces $\eta(t_1,t_2)$ in a 9 dimensional space is partially visualized on the right with 3 selected entries. }
\label{fig:embed_1}
\end{figure}

\begin{figure}[!htb]

\minipage{0.35\textwidth}
  \includegraphics[width=\linewidth]{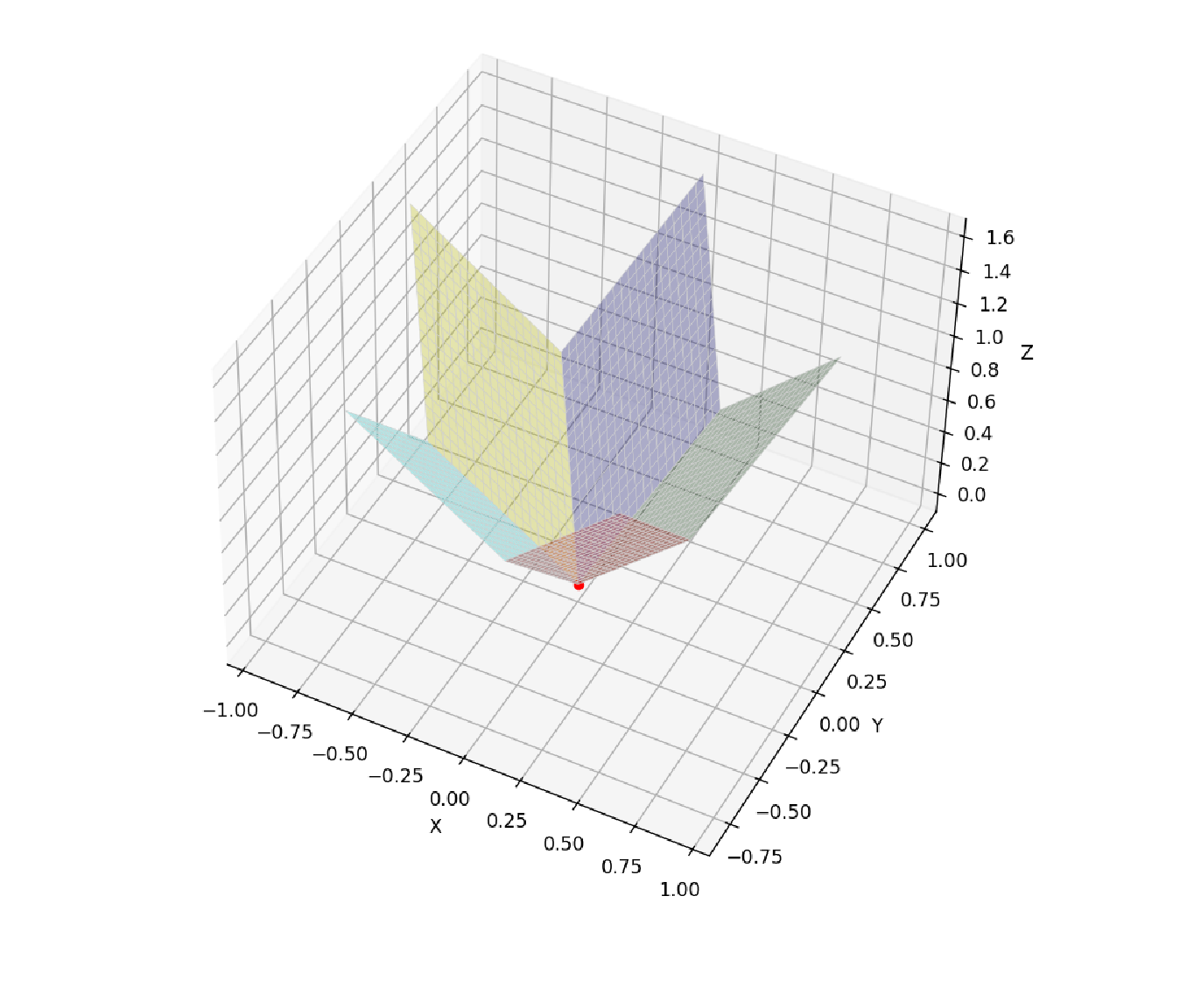}
\endminipage\hfill
\minipage{0.18\textwidth}
  \includegraphics[width=\linewidth]{arrow.eps}
\endminipage\hfill
\minipage{0.45\textwidth}%
  \includegraphics[width=\linewidth]{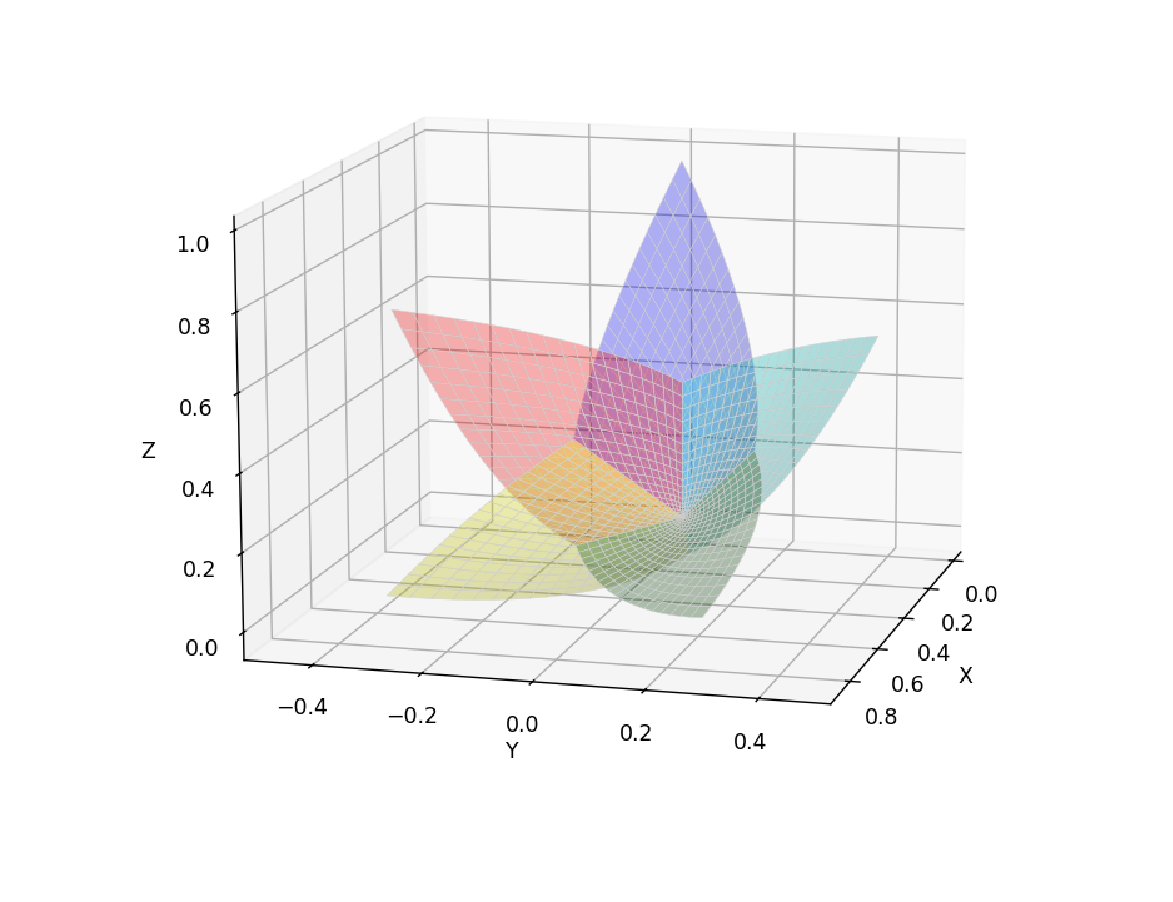}
\endminipage
\caption{Pictorial description of the embedding of 5 surfaces in a $\mathcal{Q}_5$ region. The resulting surfaces in a 9 dimensional space is partially visualized on the right with 3 selected entries.}
\label{fig:embed_2}
\end{figure}

Since the embedding $j(x)=xx^T$, $R^3 \to Sym(3)$ put the surface in a higher dimensional space of 9 dimensions (6 after removing redundant ones,), it is impossible to provide a complete visualization. However, one can provide a partial, 3-D visualization of the embedded images by selecting 3 entries out of any element, $A \in Sym(3)$. In the following displays, We selected $a_{11}$, $a_{21}$ and $a_{22}$, for a given data point $A =\begin{bmatrix}
    a_{11} & a_{12} & a_{13} \\
    a_{12} & a_{22} & a_{23} \\
    a_{13} &  a_{23} & a_{33}
    \end{bmatrix}$.
In figure \ref{fig:embed_1}, The embedded images of a single triangle and rhombus is shown. Intuitively, the embedding produces curved surfaces in a high dimensional space, due to the quadratic terms that parameterize the resulting surfaces. In figure \ref{fig:embed_2}, we display the relevant embedded images of 5 triangles/rhombi in a quadrant (or sometimes named as a Q5 region, described earlier.)

The images in figure.\ref{fig:embed_2} show that the overall spatial orientation of the surfaces is preserved by the embedding. Although the embedded surfaces from each individual triangle/rhombus seem to have more overlap compared to that of the original space, we know that theoretically the overall topology of the embedded surface must not change under the embedding map, which is a homeomorphism. Therefore, this visual character is likely an artifact caused by visualization in only a lower dimensional subspace. By the same reasoning, the neighborhood structure of each individual surface shouldn't change either, although the embedded surfaces seem more convoluted, forming a star-like pattern instead of the cone-like pattern in the original space. For completeness, we included the visualization of 3D marginals, for all 15 embedded rhombi in the appendix; some of them are degenerated .

Finally, since our goal is to analyze data supported on the whole $\mathcal{T}_4$, we also provide a visualization for all 15 rhombi selected for analysis, as a preparation for the following section.




\subsection{Calculation of the Projection $P_j\overline{j(X)}$ with Analytical Solution}

Without loss of generality, we consider the projection of the sample mean in the embedded ambient space on to a given rhombus. To achieve this, We first try solving a simpler problem, by projecting the Euclidean sample mean in the embedded space $\hat{\mu}=\overline{j(X)} \in sym(3,3)$, on to the embedded manifold of  $R^3$, i.e, in this case, $j(\mathcal{M})=j(R^3)=\{yy^T, y\in R^3\}$. This is essentially the original Venrones-Whitney extrinsic mean, without the unit norm assumption.

Let $\hat{\mu}=\overline{j(X)}$, and assuming the matrix $u$ has a unique largest eigenvalue $\lambda_N$, with the eigenvector $v_N$, then:

\[
\mbox{min}( Tr(yy^T-\hat{\mu})(yy^T-\hat{\mu})^T))=\mbox{min}( Tr(yy^T-\hat{\mu})(yy^T-\hat{\mu}))
\]
\[
=\mbox{min}(Tr(yy^Tyy^T)-2Tr(yy^T\hat{\mu}))=\mbox{min}(Tr(y^Tyy^Ty)-2Tr(y^T\hat{\mu} y))
\]
\[
=\mbox{min}(||y||^4-2y^T\hat{\mu} y) \geq ||y||^4-2\lambda_N||y||^2
\]

let $\psi=||y||^2 $, the last quantity is a quadratic equation with respect to $\psi$:
\[
\psi^2-2\lambda_N\psi
\]

Which has a unique minimum at $\psi=\lambda_N$. Therefore, the optimal solution for $y$ is $\sqrt{\lambda_N}v_N$.

However, this is not enough to calculate the projection on to (the embedded surface) of SPED, in which case, $j(\mathcal{M})=j(R^3 \cap SPED)=j(SPED)=\{yy^T, y\in SPED\}$ . Although we eliminated the assumption of $||y||=1$, such as in the case of projective sphere, we need to also constrain $y$ to be a point on the SPED's surface.

\begin{figure}[!h]
\centering
\includegraphics[scale=0.3]{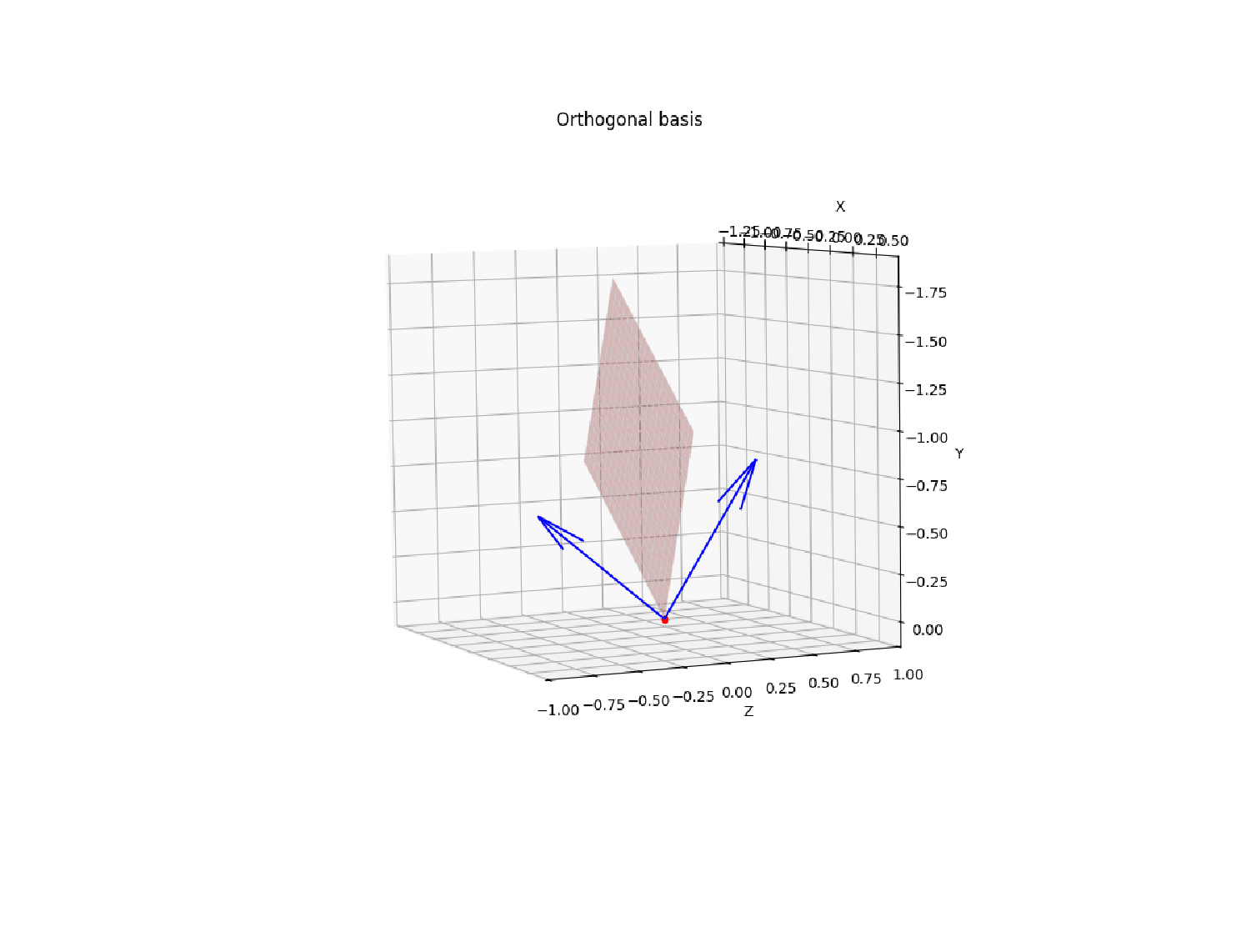}
\caption{\scriptsize An orthogonal basis for representing points on a given rhombus.}
\label{fig:bor}
\end{figure}

To achieve this, we use the same method as above, by essentially reducing the problem to a quadratic optimization problem in the form of eq.\eqref{eq:quad}, where $\psi=||y||^2 $, and k is the maximum of $y^T\hat{\mu} y$, restricting $||y||=1$, and $y$ being a point on a certain rhombus.
\begin{equation}
   \psi^2-2k\psi \label{eq:quad}
\end{equation}
Because $y$ now lives on a 2-D subspace, finding $k$ is now an optimization problem. To this end, we first parametrize a point on a rhombus, by introducing a pair of orthonormal vectors $e_1$, $e_2$, as in figure \ref{fig:bor}.
We now have $y=\xi_1e_1+\xi_2e_2$. As one can easily observe, for $y$ to be supported on the rhombus, we will need that $\xi_1\in I_1 \subset [0,1)$ and $\xi_2\in I_2\subset[0,1)$.
We would like to further express $y$ in the coordinate system spanned by the eigenvectors of $\mu$. Writing the pairs of eigenvalues/eigenvectors of $\mu$ as  $(\lambda_i,v_i)$, we have $y=\sum_1^3c_iv_i$, where $c_i=<y,v_i>=<\xi_1e_1+\xi_2e_2,v_i>=\xi_1<e_1,v_i>+\xi_2<e_2,v_i>$.
Now we can re-write $y^T\hat{\mu} y$ as:

\begin{align}
y^T\hat{\mu} y &= (\sum_1^3c_iv_i^T)(\sum_1^3\lambda_ic_iv_i) \nonumber\\
         &= \sum_1^3\lambda_ic_i^2 \nonumber\\
         &= \xi_1^2(\sum_1^3\lambda_i<e_1,v_i>^2) +\xi_2^2(\sum_1^3\lambda_i<e_2,v_i>^2) + \nonumber\\
           &  2\xi_1\xi_2(\sum_1^3\lambda_i<e_1,v_i><e_2,v_i>). \nonumber\\
        &=\xi_1^2r_1 + \xi_2^2r_2+2\xi_1\xi_2r_3 \label{eq:temp} \\
        & r_1=\sum_1^3\lambda_i<e_1,v_i>^2 \nonumber\\
        & r_2=\sum_1^3\lambda_i<e_2,v_i>^2 \nonumber\\
        & r_3=\sum_1^3\lambda_i<e_1,v_i><e_2,v_i>
\end{align}

As described earlier, in this first stage, we focus on the case where $||y||=1$. Therefore, we can do a substitution here using: $x=\xi_1^2$, and $1-x=\xi_2^2$, $x\in I_x \subset [0,1)$.

Expression \eqref{eq:temp} is now:
\begin{equation}
k(x)=xr_1+(1-x)r_2+2\sqrt{x(1-x)}r_3 \label{eq:kx}
\end{equation}

The shape of \eqref{eq:kx} depends on the sign of $r_3$. Specifically, (i) when $r_3$ is positive, \eqref{eq:kx} is a concave function, therefore, if there is a unique maximum $x^*$ in $I_x$, then we are done. If $x^*$ is outside of $I_x$, then the solution should be on the boundary of $I_x$.  (ii) When $r_3 \leq 0$, the function is either linear or convex. The maximum(s) is therefore always on the boundary of $I_x$. (See a visualization in the next section).

Since we found in our experiments that $r_3$ is positive for the best surface to project $\mu$ onto, we will start with the first case (i), where \eqref{eq:kx} is concave:

The maximization of $k(x)$ requires nothing but taking the derivatives of \eqref{eq:kx} and setting them to 0. After rearranging terms and solving a quadratic problem, we find that :
\[
x^*=\frac{1}{2}(1+\mbox{sign}(r_1-r_2)\sqrt{\frac{\beta}{\beta+4}})
\]
, where $\beta=(\frac{r_1-r_2}{r_3})^2$.
This means that $(\xi_1^*)^2=x^*$ will be larger then $\frac{1}{2}$ if the contribution of the first basis $e_1$ is bigger than $e_2$ , i.e., $r_1 > r_2$.
$k_{max}$ is therefore $k(x^*)$. With similar reasoning as before, the optimal norm of $y$ for \eqref{eq:quad}, where $k$ is replaced by $k_{max}$, is $\sqrt{k_{max}}$.

Finally, the optimal $y$ is then:
\[
y^*=(\sqrt{x^*}e_1+\sqrt{1-x^*}e_2)\sqrt{k(x^*)}
\]

For completeness, we can also solve for the optimal $y$ on any boundary of a rhombus. Let
$y=te_a$, $t \in [0,1]$, where $e_a$ is a vertex of the rhombus, the problem is simply minimizing $\psi^2-2k\psi$,
where $k=\sum_1^3\lambda_i<e_a,v_i>^2$ and $\psi=t^2$.
The optimal $y$ is therefore: $\sqrt{\sum_1^3\lambda_i<e_a,v_i>^2}e_a$.

In summary, determining the projection for this problem involves two steps:
Step 1: solve
\[
k_{max}= y^T\hat{\mu} y, ||y||=1
\]
Step 2: solve for $||y||$ while minimizing:
\[
\psi^2-2k_{max}\psi
\]
where $\psi=||y||^2 $.

The correctness of the solution will be verified through experiments with the real yeast dataset, as we will demonstrate in the next section.
\subsubsection{A Closer Look at The Solution}
In this section, we expand on the characterization of the solution in the previous section. One can see that out of the two steps, the first step is more fundamental, since it determined the direction of the optimal $y$.  Recall that the most important coefficient is $r3$ in \eqref{eq:kx}. In both simulations and the real data experiment, the first eigenvalue $\lambda_1$ is usually dominantly large (5 to 10 times of $\lambda_2$ and $\lambda_3$), and the influence of the first eigenvector is usually the largest:
\[
r_3=\sum_1^3\lambda_i<e_1,v_i><e_2,v_i>  \approx \lambda_1<e_1,v_1><e_2,v_1>
\]

\begin{figure}[!h]
\centering
\includegraphics[scale=0.5]{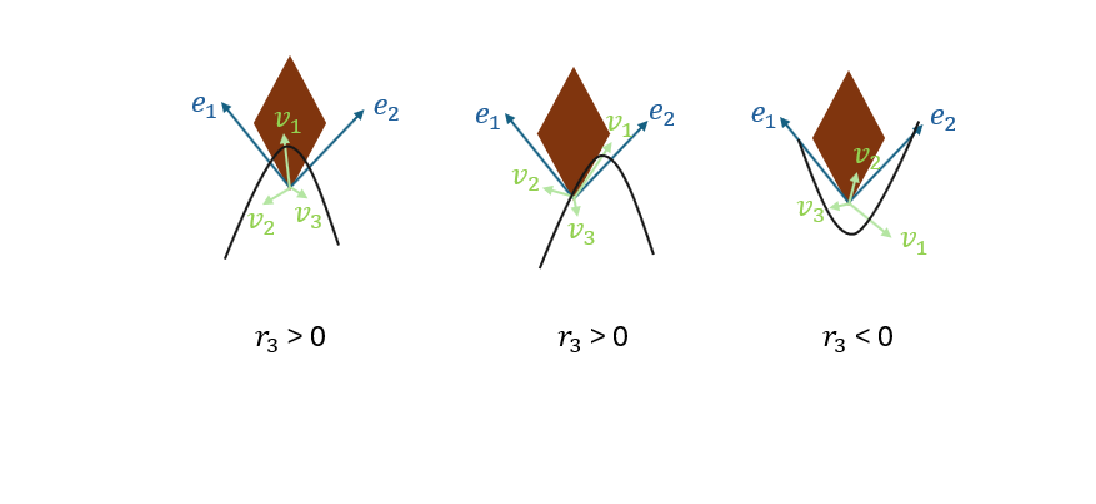}
\caption{\scriptsize Characterization of the projection onto individual rhombi with different conditions.}
\label{fig:solu}
\end{figure}

Therefore, when can roughly categorize the projection into the three types, as depicted in Fig \ref{fig:solu}.

Therefore, the rhombi that are close to $v_1$ will have a chance to have a projection in the interior. For surfaces far away from $v_1$ , the projection is always on a boundary.

\subsubsection{The Complete Formula for the Projection}
The full formula for the projection is then:

\[P_j(\mu)=y^*(y^*)^T=
k(x^*)(x^*e_1e_1^T+2\sqrt{x^*(1-x^*)}e_1e_2^T+(1-x^*)e_2e_2^T)
\]

For the later section 3.3, in order to calculate the confidence region of the mean tree, one would need to calculate the differentials of the projection.
One can glean that if one wishes to calculate the differentials of the projection at $\mu$, i.e., of $d_\mu P_j$, the most important part  is to calculate the differentials of the coefficients $r_1,r_2$ and $r_3$, at $\hat{\mu}$. To succeed in this task, one needs to first find the differentials of all the eigenvalues $d_{\mu} \lambda_i$ and eigenvectors $d_{\mu} v_i$ of $\mu$ itself.
This could be a daunting task. As described later, we applied an automated differentiation tool and obtained the differentials without manual computation.

\subsubsection{Finding the Best Rhombus/Stratum}
Assume a unique extrinsic sample mean exists, which is equivalent to saying that there is a unique rhombus $l\in \{1,2,...,15\}$ whose embedded surface has the minimum projection loss. The extrinsic sample mean is then the inverse of such projection.

\[
\mu_{j,E,n} = j^{-1}(P_j^l\overline{j(X)})
\]

Where, $P_j^l\overline{j(X)}$ denotes the projection of $\overline{j(X)}$ onto the l-th rhombus.

\section{Application on a Yeast Genome Dataset}
\label{sec:others}
\subsection{Data Description}
The dataset we used in this section comes from \cite{1011yeast}, which contains 1011 Saccharomyces cerevisiae isolates. The researchers picked this specific strain of yeast due to its rich genetic diversity and complex population structure. In the original dataset, both the genotypes (gene sequence) and phenotyes (traits) are present.  Our analysis only used the genotype dataset, which comes with pre-aligned sequences, and a readily available distance matrix. According to the paper, the template sequence used for alignment is the S. cerevisiae S288C reference genome, where the genomic distances are calculated with the 'bionj' algorithm in R. This dataset is publicly available here: {http://1002genomes.u-strasbg.fr/files/}.

Since the preprocessing for phylogeny analysis is not trivial and can influence the results greatly, we dedicate the next section to the preprocessing pipeline we adopted.
\subsection{Data Preprocessing}
\subsubsection{Tree Building Procedures}
\begin{figure}[h]
\begin{center}
\includegraphics[scale = 0.35]{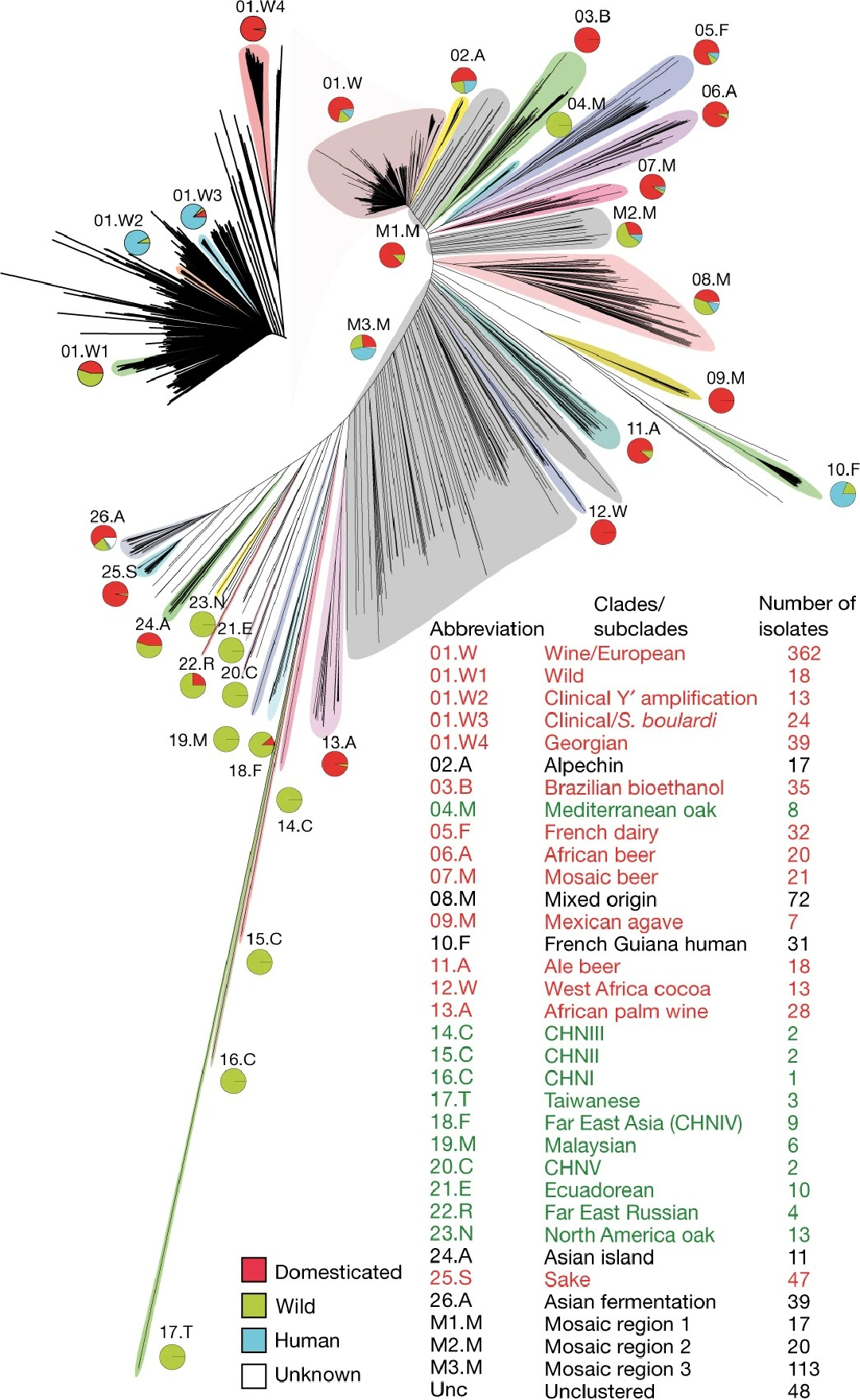}
\end{center}
\caption{Phylogenetic tree of the 1,011 strains of Saccharomyces cerevisiae, taken from (Peter et al., 2018). The tree is built with neighbor-joining method according to the original authors.}%
\label{fig:1011yeast}%
\end{figure}

In \cite{1011yeast}, the 1011 strains have already been organized into 26 clades (clusters) and 3 mosaic regions.  A phylogenetic tree involving all the strains is available in the original paper, as displayed below (figure. \ref{fig:1011yeast}).

According to the authors, this tree was built using the neigbor-joining method, which is adopted in our experiment for consistency. The neighbor-joining method is originated in \cite{Saitou1987TheNM}, which has some major advantages over traditional clustering methods, such as hierarchical clustering, with one of them being that the cluster members are not assumed to be equidistant to the cluster centroid. The algorithm also performs an optimization which aims to "minimize the evolution needed" to explain difference among objects \cite{hall2004phylogenetic}.  Literature suggests that phylogenetic trees built with the neighbor joining method are likely to be more topologically accurate \cite{mihaescu2009neighbor}. For our experiment, we first pick 4 clades/population from the yeasts family considered in \cite{1011yeast}, and then we sample without replacement 1 strain from each individual clade. The tree built with the resulting 4 strain samples constitute one sample tree, and we will repeat this process to create more sample trees. Based on the genomic distances provided by the original authors, we used the '\textbf{nj}' method from the \textbf{scikit-bio} package in python to built the neighbor-joining trees, and  the \textbf{biopython} package to compare tree topologies and extract the length of interior edges.
\subsubsection{Rooting a Tree}
Clustering methods such as the neighbor-joining method inherently produces unrooted trees.
However, in order to equip ourselves with a dataset on $\mathcal{T}_4$, we need to provide a root to the trees (see another supplemetary reference in \cite{hall2004phylogenetic}, showing that there are only 3 types of unrooted binary trees with 4 leaves instead of 15). Ideally, to correctly root the yeast trees require prior knowledge of the real ancestral origin of the yeast strains in consideration, which is still in debate at the moment. Although historical evidence suggested that yeasts around the world may share a single origin in China, consensus has not been reached on whether this is a fact or still a hypothesis at large. In fact, the paper itself \cite{1011yeast} suggested a single 'out-of-China' origin. For our analysis, we used one Chinese strain as the root for conveniences (this does not diminish the value of our analysis, since one does not always have to find the evolutionary ancestor to convert an unrooted tree to a rooted tree. One can instead pick a group that is sufficiently distant from the leaf groups of consideration. This practice is called 'out-grouping', see \cite{hall2004phylogenetic} for more details). In this sense, the stem distance (distance from the root to the first interior node) is interpreted as the relative distance of the 4 leaf populations to the out-group, instead of the evolutionary distance from their common ancestor. Although the stem distance was not included in our analysis, rooting the trees are necessary for creating a interior node to 'dangle'/'hang' the trees from, such that they can be placed on BHV's $\mathcal{T}_4$. If the stem distance is relevant for the analysis, practitioners can consider a space of $\mathcal{R} \times \mathcal{T}_4$, which is a natural extension for future works.

\subsubsection{Picking Populations}
\begin{figure}[h]
\begin{center}
\includegraphics[scale = 0.6]{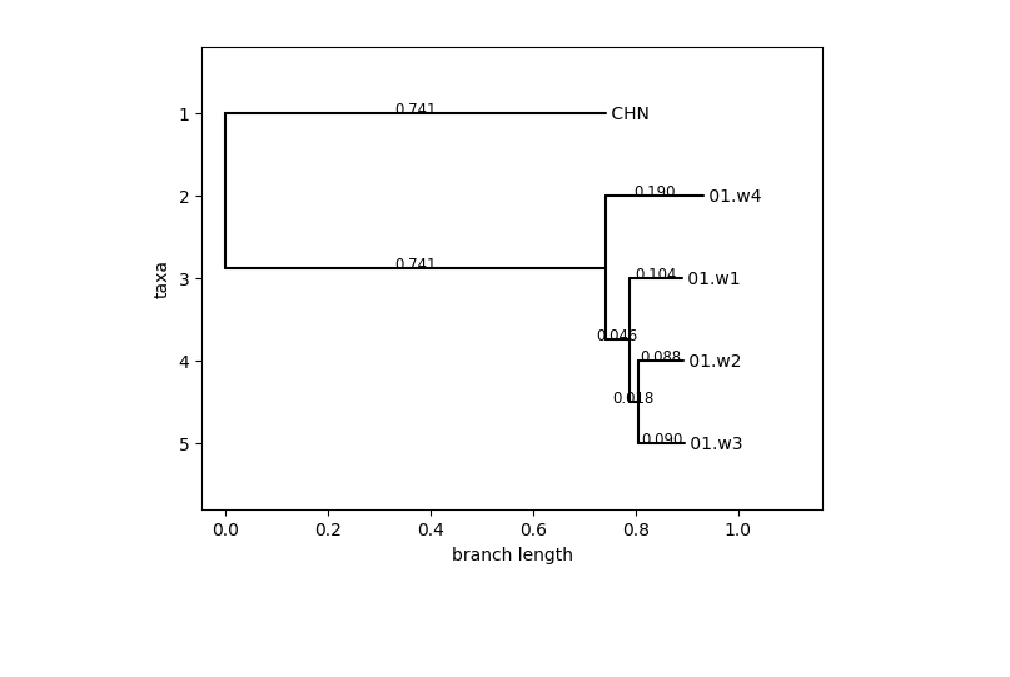}
\end{center}
\caption{A phylogenetic tree of 4 Wine strains taken from the 4 sub-clades within the European-wine clade, respectively. The tree is built with the neighbor-joining method, consistent with the method in (Peter et al., 2018). The apparent difference in appearance from this tree and the previous figure is only due to configurations of the displaying mode. }%
\label{fig:winetree}%
\end{figure}

\begin{figure}[!htb]
\minipage{0.33\textwidth}
  \includegraphics[width=\linewidth]{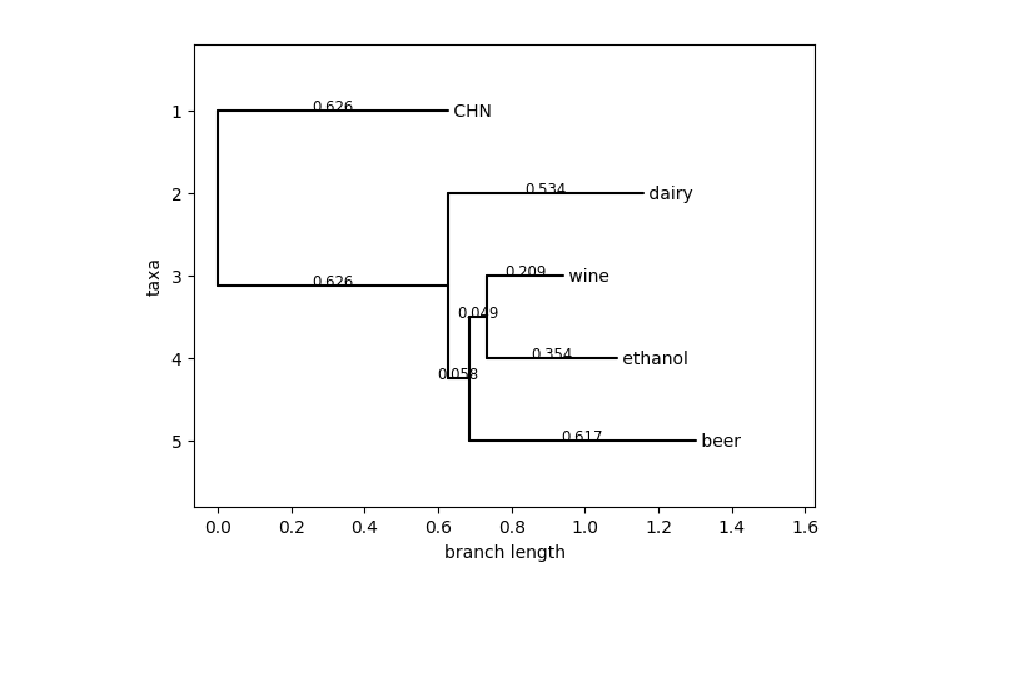}
\endminipage\hfill
\minipage{0.33\textwidth}
  \includegraphics[width=\linewidth]{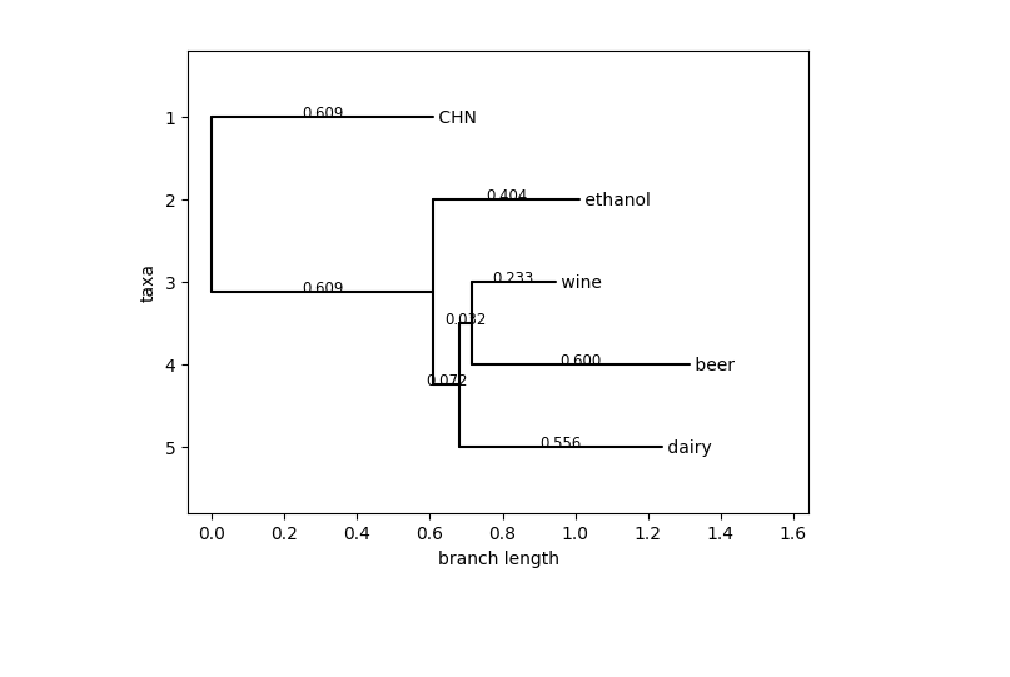}
\endminipage\hfill
\minipage{0.33\textwidth}%
  \includegraphics[width=\linewidth]{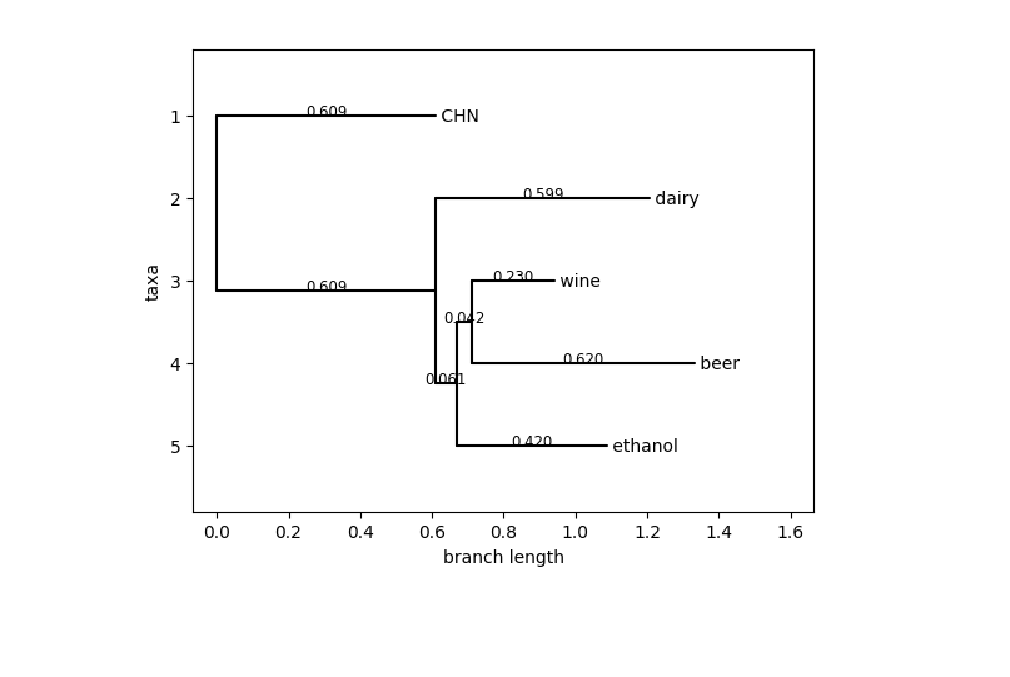}
\endminipage\hfill
\minipage{0.33\textwidth}
  \includegraphics[width=\linewidth]{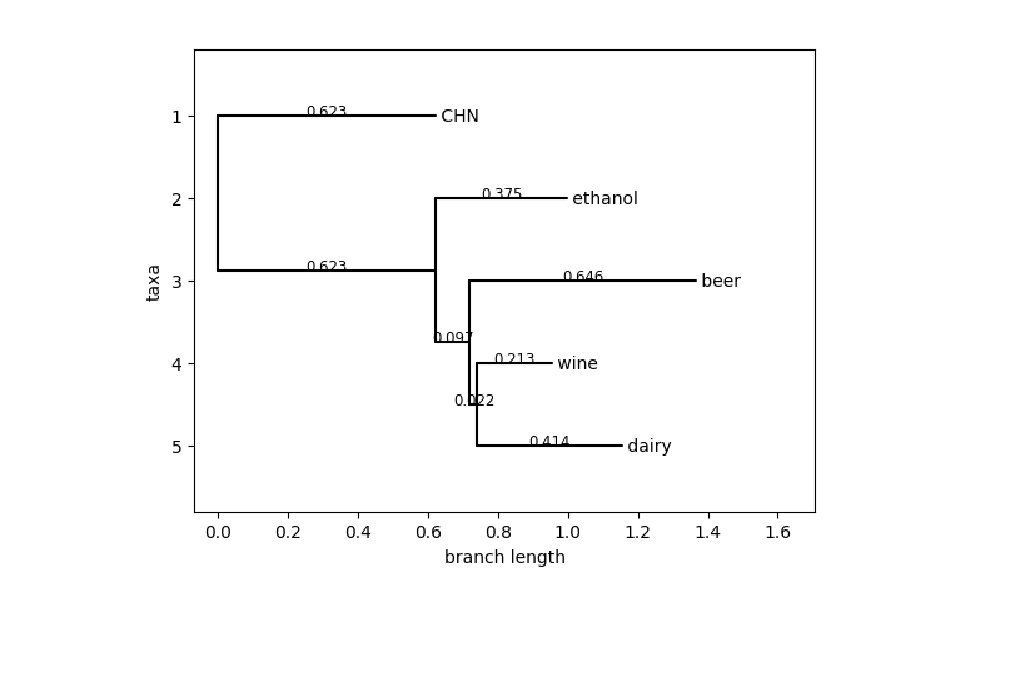}
\endminipage\hfill
\minipage{0.33\textwidth}
  \includegraphics[width=\linewidth]{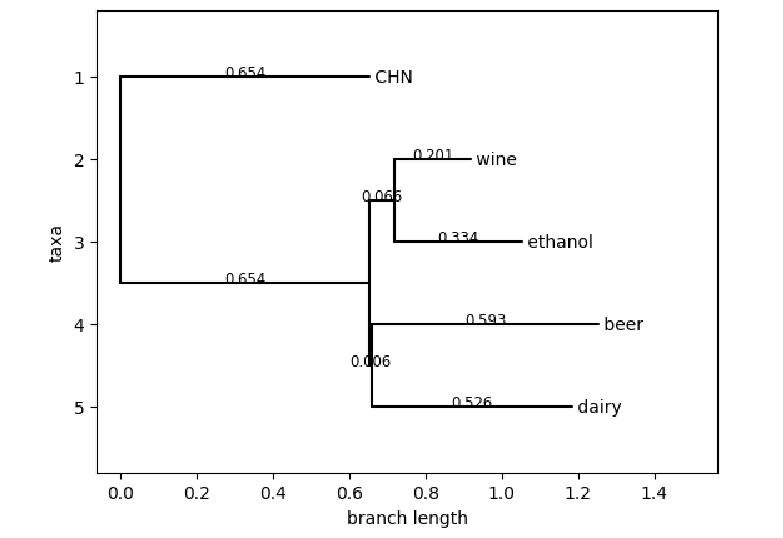}
\endminipage\hfill
\minipage{0.33\textwidth}
  \includegraphics[width=\linewidth]{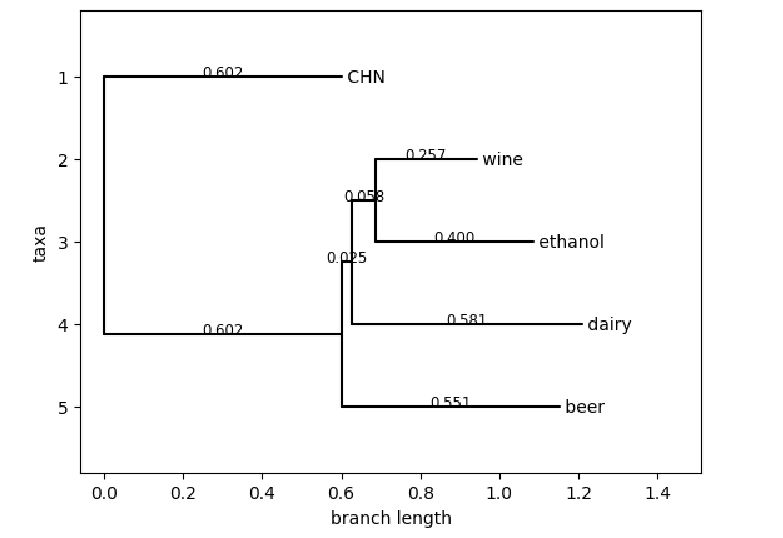}
\endminipage
\caption{6 different tree types built from 4 more distance strains of yeasts (1:Wine/European, 2:A African beer, 3: Brazilian bio-ethanol, 4: French dairy), rooted with a Chinese yeast strain. The trees are sorted in decreasing order based on their frequencies.}
\label{fig:yeast trees}
\end{figure}

When picking different populations to build a phylogenetic tree, we would want populations that are distinct from each other, such that they can be represented as distinct leaf nodes on the tree. However, we also want the strains within each population to have roughly the same genetic lineage. Since this later assumption is likely to be violated in the mosaic groups, we excluded the 3 mosaic populations in \cite{1011yeast} and only worked with the 26 clearly identified clades. Originally, we picked the four sub-clades within the European-wine population, namely, 01.W1-W4 within the Wine/European group 01.W. However, the resulting trees are all of the same type (figure. \ref{fig:winetree}), potentially due to the fact that these 4 clades are tightly clustered, with very little within-group variance. Therefore, we decided to expand our search to other clades with more dispersion. Our first attempt using the other 4 groups resulted in much more variability, which includes the following 4 clades: 01.W Wine/European (n=362), 03.B Brazilian bioethanol (n=35), 05.F French dairy (n=32) and 06.A African beer (n=20). The experiment produced 6 types of trees, as displayed in figure \ref{fig:yeast trees}. Finally, it is worth noting that due to the fact that we have built tree with sample strains from each clade without replacement, the total sample size (number of sample trees) is the minimum of sample sizes of the 4 clades, which is 20, from African beer. The trees in figure. \ref{fig:yeast trees} have decreasing frequencies, respectively being 12, 3, 2, 1, 1, and 1.

\clearpage

Finally, we visualize the sample trees on the 15 selected rhombi. Due to the scale of the interior edges, the original data are very close to the center of the SPED. The distribution can be seen more clearly after magnification by a factor of 5. As one can see, the 20 sample trees concentrate on a few neighboring rhombi, with slight dispersions. The most dominant tree type is supported on the 15-th, brown surface, highlighted with the trace of its boundary (the indexing order of the surface is arbitrary).

\begin{figure}[!htb]
\minipage{0.5\textwidth}
  \includegraphics[width=\linewidth]{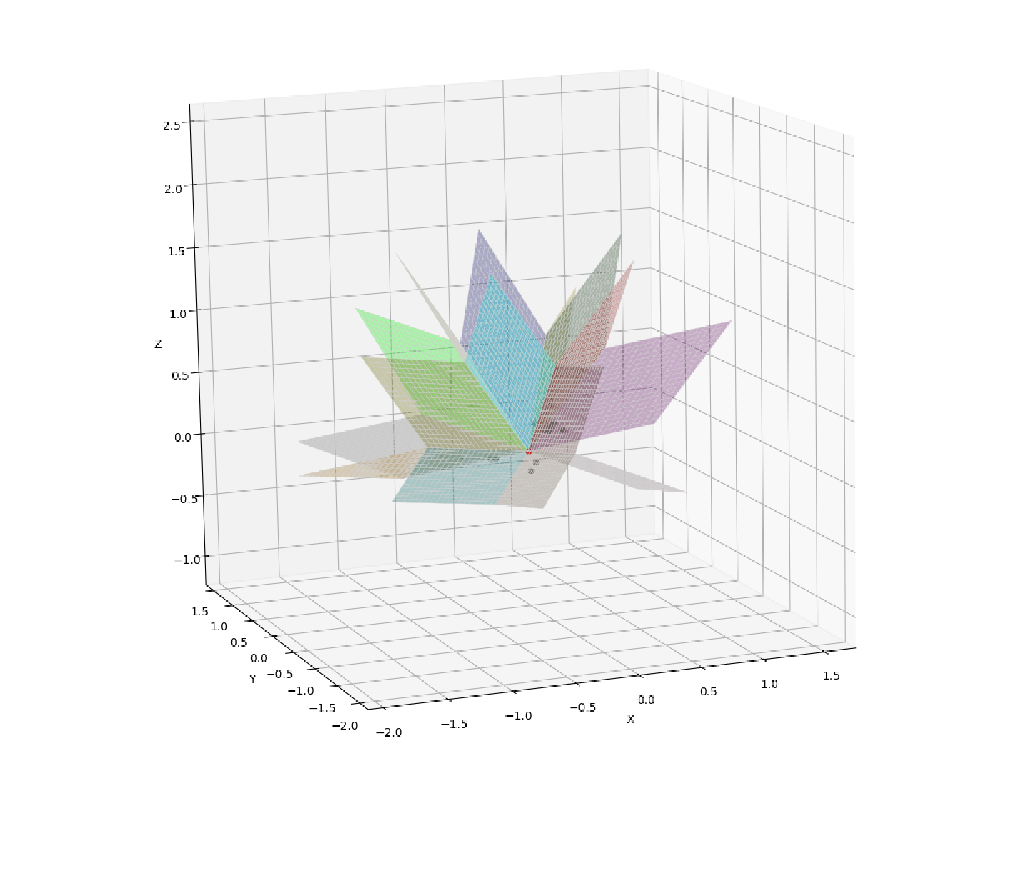}
\endminipage\hfill
\minipage{0.5\textwidth}
  \includegraphics[width=\linewidth]{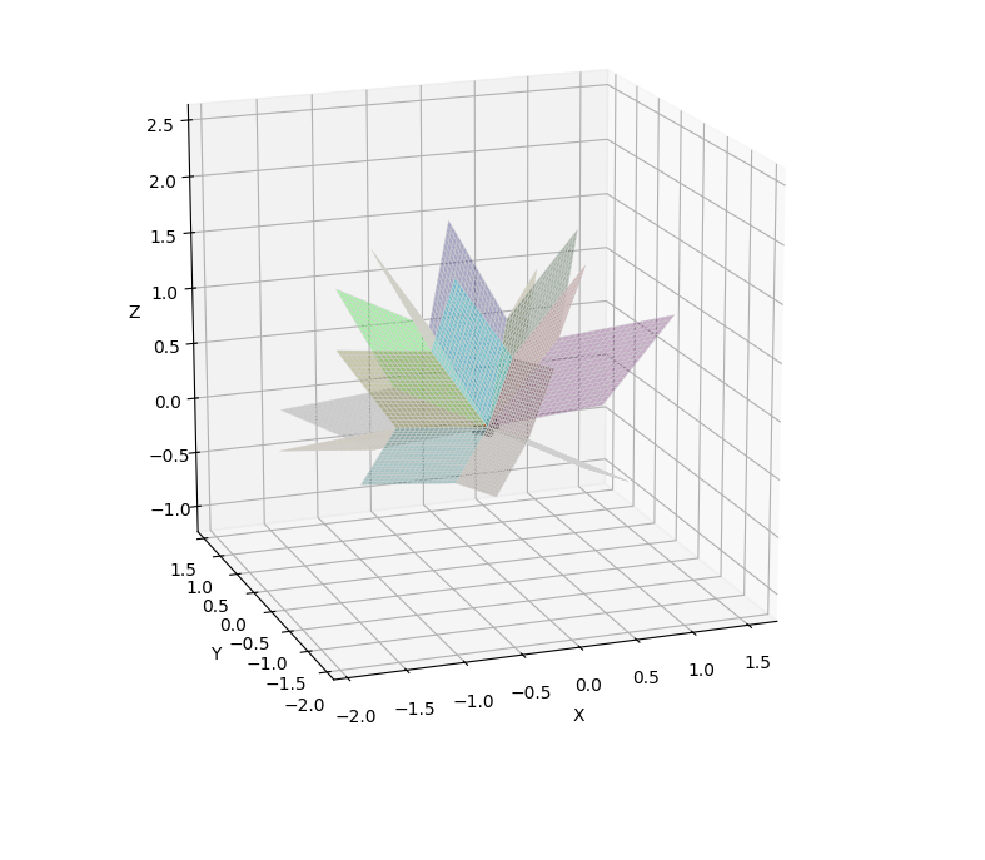}
\endminipage
\caption{The 20 sample yeast trees displayed on SPED. the original data are very close to the origin (Right). The distribution can be seen more clearly after magnification by a factor of 5 (Left), where the stratum that has the most samples (12 data points) is highlighted.}
\label{fig:yeast trees on SPED}
\end{figure}
\clearpage

\subsection{Inference of the Mean Yeast Tree}
With the data at hand, we first calculate the extrinsic sample mean via the analytical solution described in section 4.4.3. Specifically, the analytical solution was applied to each individual rhombus to find the optimal projection of $\overline{j(X)}$ on its embedded surface. We then calculate the pre-image of the best projection as the extrinsic sample mean $\mu_{j,E,n}$. The extrinsic sample mean turns out to be supported exactly on the rhombus/stratum where that contains the majority of the data, i.e. the 15-th (brown) surface in figure\ref{fig:yeast trees on SPED}.

Furthermore, we bootstrapped the original samples 1000 times and found that the bootstrapped extrinsic sample mean concentrates on the 15-th surface, in the sense that 999 out of the 1000 extrinsic sample mean remain on the 15-th surface, which presents strong evidence that the extrinsic sample mean is unique, and the distribution is non-focal, with non-sticky extrinsic sample means. The inference problem is then reduced to that of a manifold. Section 3.3.1-3.3.3 outlined the relevant theories from \cite{patrangenaru2016nonparametric}.
\subsubsection{Normal Approximation}
Recall the theorems of extrinsic CLT on manifold, as reviewed in Chapter 2.5 of \cite{patrangenaru2026nonparametric}, once an adapted frame is built for $P_j(\overline{j(X)})$, the tangent element on the adapted frame can be used to construct confidence intervals for the extrinsic mean in various ways.
From normal asymptotics:
\[
n \|S_{j,E,n}^{-{1\over 2}} tan_{P_j(\overline{j(X)})}
(P_j(\overline{j(X)}) -P_j(\mu))\|^2
\]
converges weakly to  $\chi^2_2.$

For the calculation of $S_{j,E,n}$ , as in eq (5.48) in \cite{patrangenaru2016nonparametric}, we adopted an (exact) auto-differentiation tool in a Python library called Pytorch for $d_{\overline{j(X)}} P_j$, which can calculate exact gradients/differentials for very complicated functions according to their computational graphs (See the official documentation here. {https://docs.pytorch.org/docs/stable/notes/autograd.html}).

\subsubsection{Non-pivotal Bootstrap}
Recall from \cite{patrangenaru2016nonparametric} that the distribution  function of $n\|tan(P_j(\overline{j(X)}) -
P_j(\mu))\|^2$ can also be approximated uniformly by the bootstrap
distribution of
$$n\|tan_{P_j(\overline{j(X)})}(P_j(\overline{j(X^*)}) -
P_j(\overline{j(X)}))\|^2$$

\subsubsection{Pivotal Bootstrap}
To produce a confidence region that is more efficient, one can use the pivotal bootstrap, which has coverage error $O_p(n^{-2})$, which is superior to both the asymptotic coverage error of order ($O_p(n^{-1})$) and non-pivotal bootstrap ($O_p(n^{-\frac{m}{m+1}})$). The differentials $d_{\overline{j(X^*)}} P_j$ are also calculated via Pytorch.

\begin{figure}[!htb]
\begin{center}
\includegraphics[scale = 0.5]{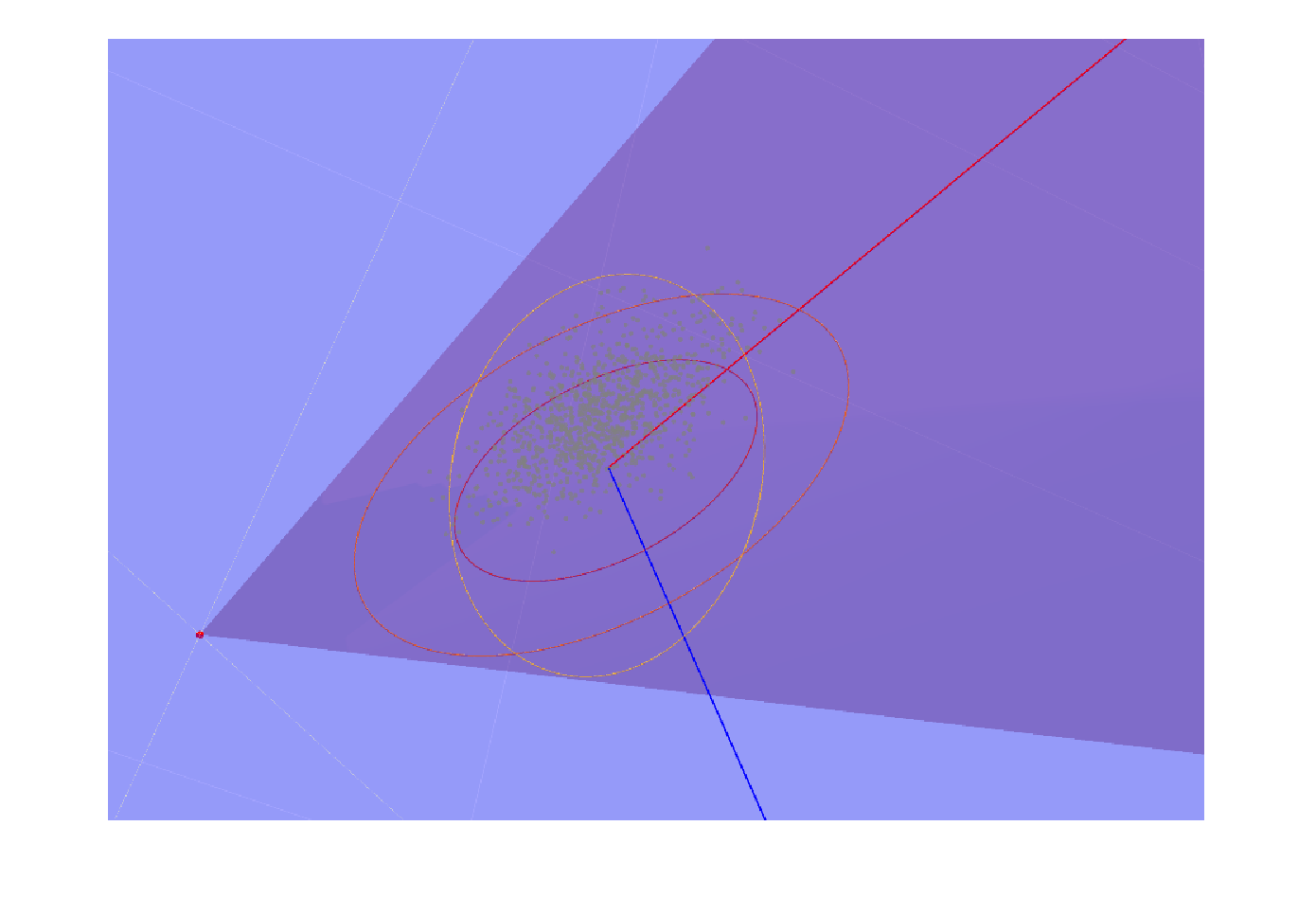}
\end{center}
\caption{The 95\% confidence regions via 1: Normal approximation (Red), 2:Non-pivotal bootstrap (Yellow), and 3: Pivotal bootstrap (orange), on the tangent plane centered at $Pj(\overline{j(X)})$. The (embedded images of the) bootstrapped extrinsic sample mean on this plane are shown in gray.}
\label{fig:confreg_3}
\end{figure}
We displayed the three resulting 95\% confidence regions on the tangent plane in Fig \ref{fig:confreg_3}. Note
that just as before, this confidence region can not be completed visualized, since it is on a plane in 9-dimensional space. Therefore, Fig \ref{fig:confreg_3} is also a partial visualization displaying 3 selected coordinates. For this reason, the orthonormal basis of the adapted frame does not
look completely orthogonal.

\subsubsection{Conclusion}
Finally, as seen in figure \ref{fig:confreg_3}, both the {parametric} and {pivotal bootstrap} intervals overcast regions that are strictly in the interior of the embedded surface of the dominant surface (Surface 15). The sample size (20) is large enough for pivotal bootstrap to be used for statistical inference. We can then reject the hypothesis that the extrinsic mean of the yeast tree lies either on the boundary or outside of the dominant surface. We want to emphasize that our procedure for building the sample phylogenetic trees does not automatically introduce correlations among the tree samples. Specifically, the sampling of genes within a specific clade (subpopulation) was without replacement, and the gene sample will be discarded once a sample tree was built. However, this result may require closer scrutiny, considering that genes within a specific clade (subpopulation) from \cite{1011yeast} may not be independently sampled.
\clearpage
\appendix
\section{Supplementary materials}
\begin{figure}[!h]

\minipage{0.35\textwidth}
  \includegraphics[width=\linewidth]{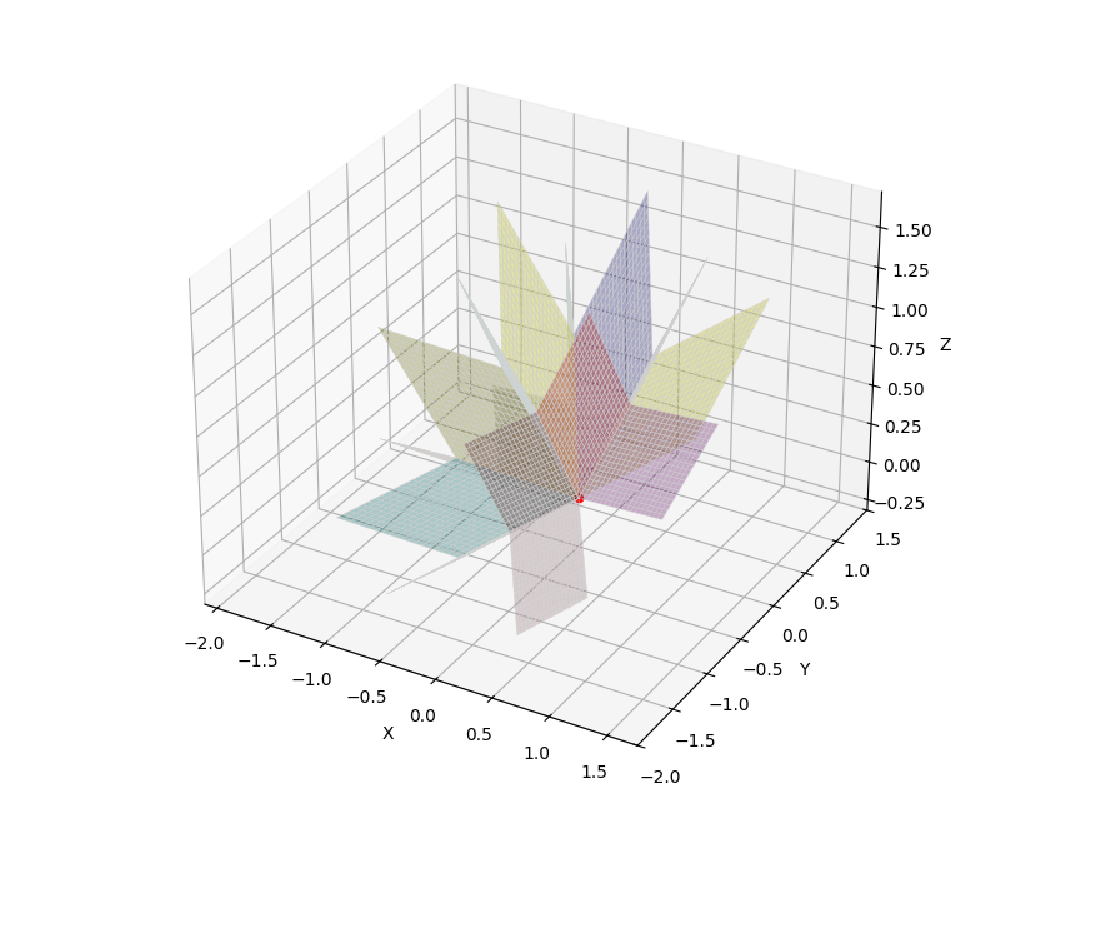}
\endminipage\hfill
\minipage{0.18\textwidth}
 \includegraphics[width=\linewidth]{arrow.eps}
\endminipage\hfill
\minipage{0.45\textwidth}%
\includegraphics[width=\linewidth]{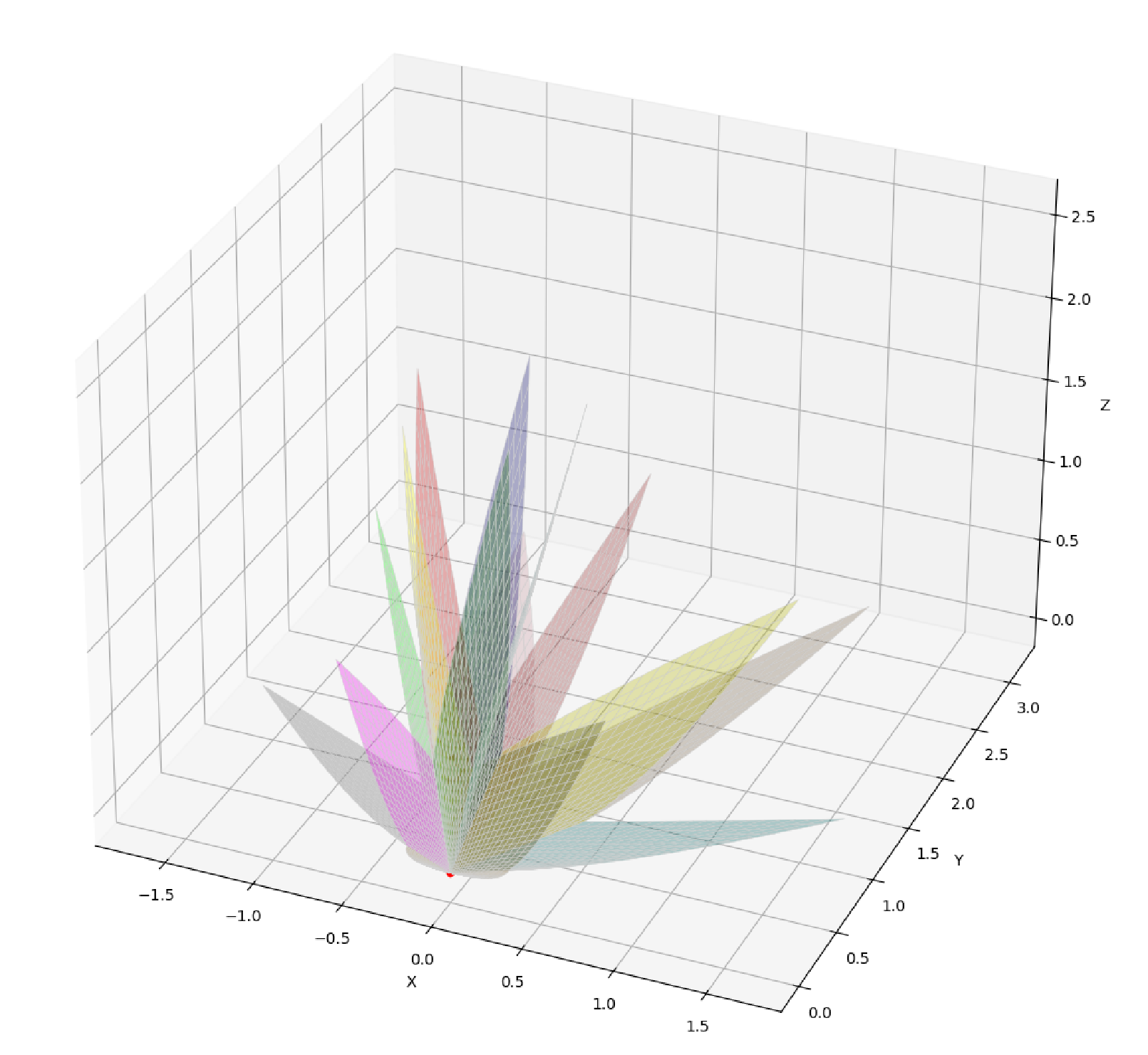}
\endminipage
\caption{Pictorial description of the embedded surfaces of 15 rhombi, which are selected as the support for our data for analysis The selected coordinates are $a_{12},a_{22},a_{33}$, respectively, from the 3 by 3 matrices.}
\label{fig:embed_4}
\end{figure}

Note that in general, a partial collection of coordinates may not be able to show all 15 surfaces. Consider a simpler case, for example, if the coordinate system is selected such that one of the a unit squares has the third coordinates $x_3=0$, the embedded surface is now:
$\eta(t_1,t_2) = t_1^2 \begin{bmatrix}
    1& 0 & 0 \\
    0 & 0 & 0 \\
    0 &  0 & 0
    \end{bmatrix}+t_2^2\begin{bmatrix}
    0& 1 & 0 \\
    0 & 0 & 0 \\
    0 &  0 & 0
    \end{bmatrix} + t_1t_2 \begin{bmatrix}
    0& 1 & 0 \\
    1 & 0 & 0 \\
    0 &  0 & 0
    \end{bmatrix} $.
Many choice of three coordinates will result in co-linear coordinates. In particular, a choice of $a_{13},a_{23},a_{33}$ will cause the resulting surface to collapse to a point in the origin.

\begin{figure}[!htb]
\minipage{0.2\textwidth}
    \includegraphics[width=\linewidth]{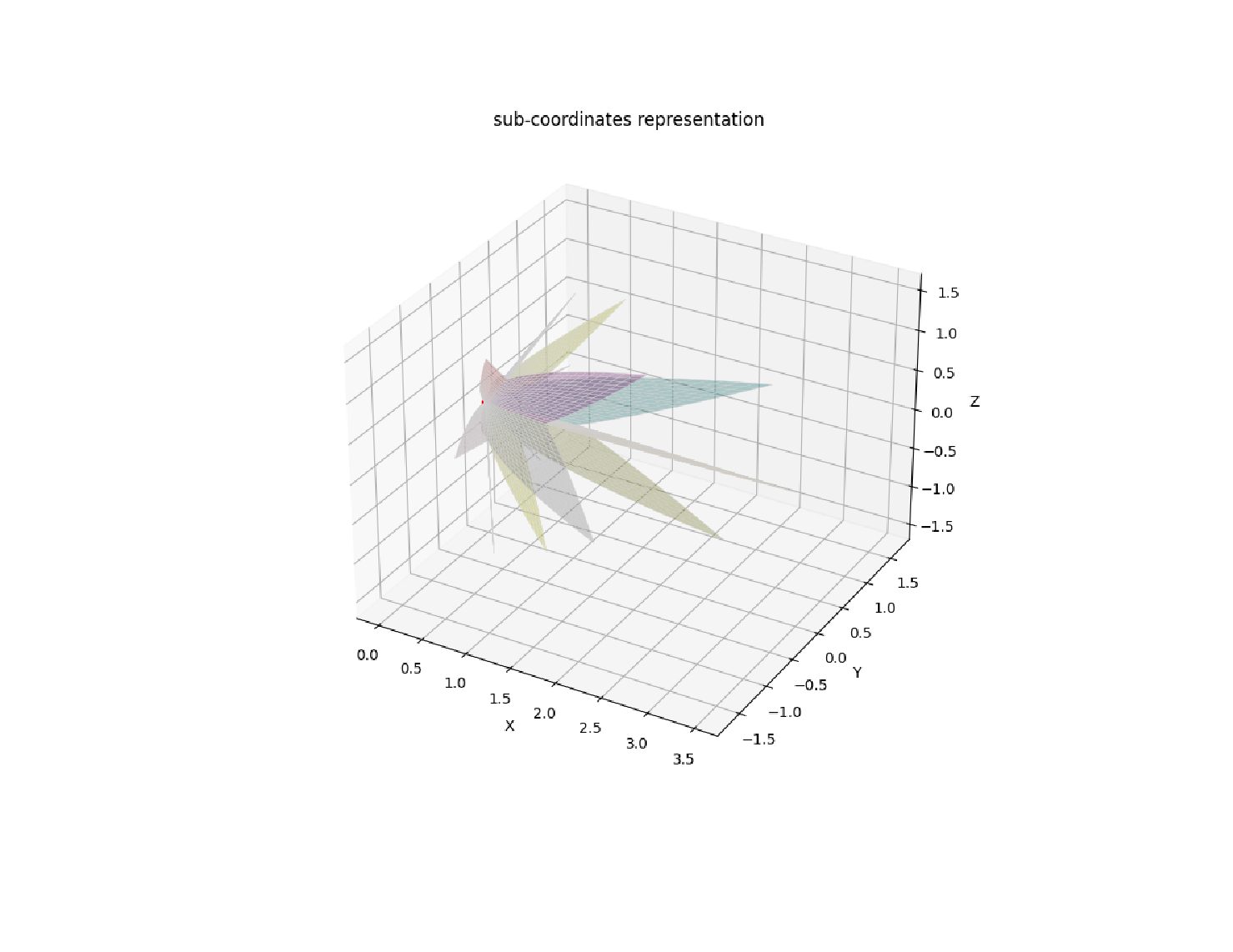}
\endminipage\hfill
\minipage{0.2\textwidth}
    \includegraphics[width=\linewidth]{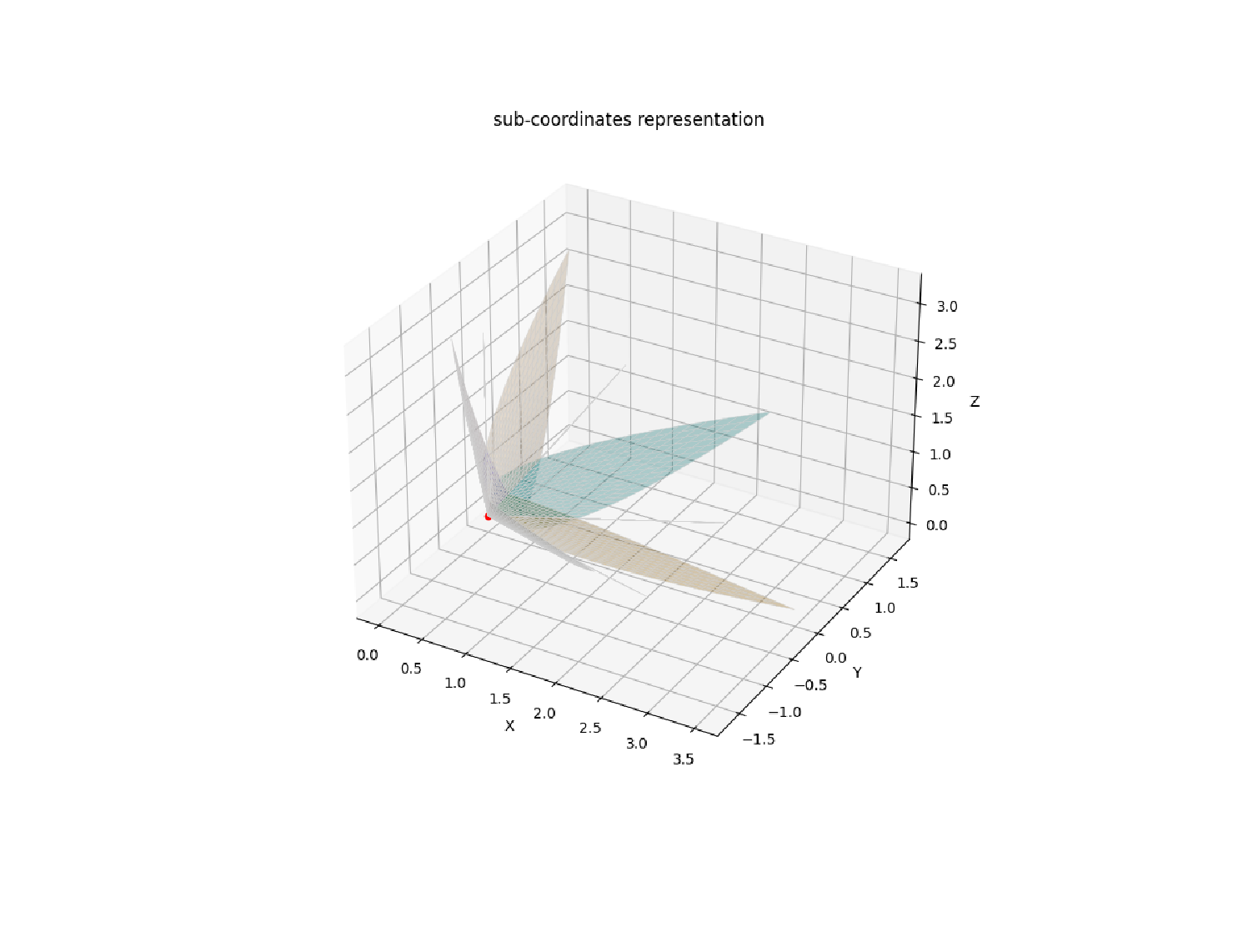}
\endminipage\hfill
\minipage{0.2\textwidth}%
    \includegraphics[width=\linewidth]{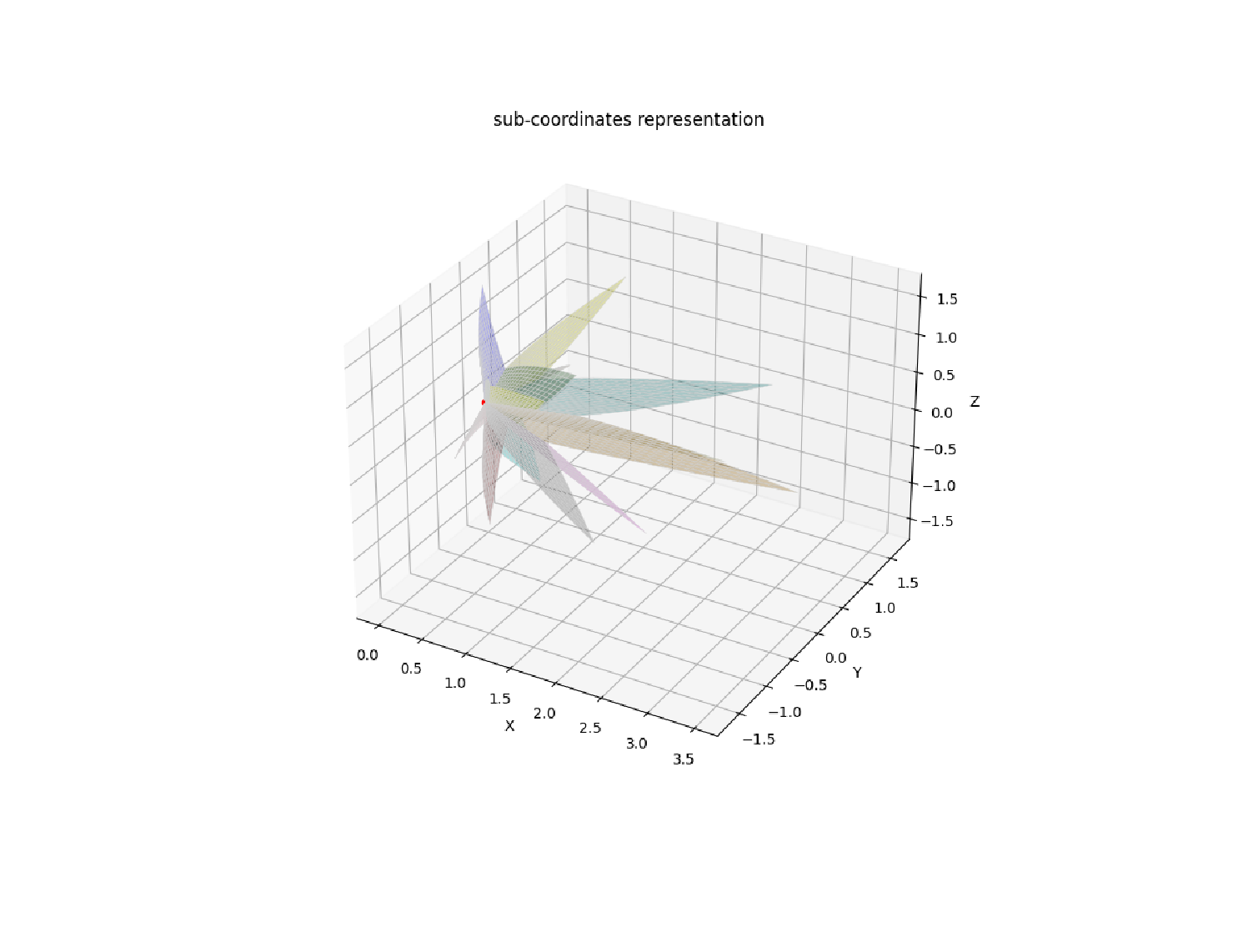}
\endminipage\hfill
\minipage{0.2\textwidth}
    \includegraphics[width=\linewidth]{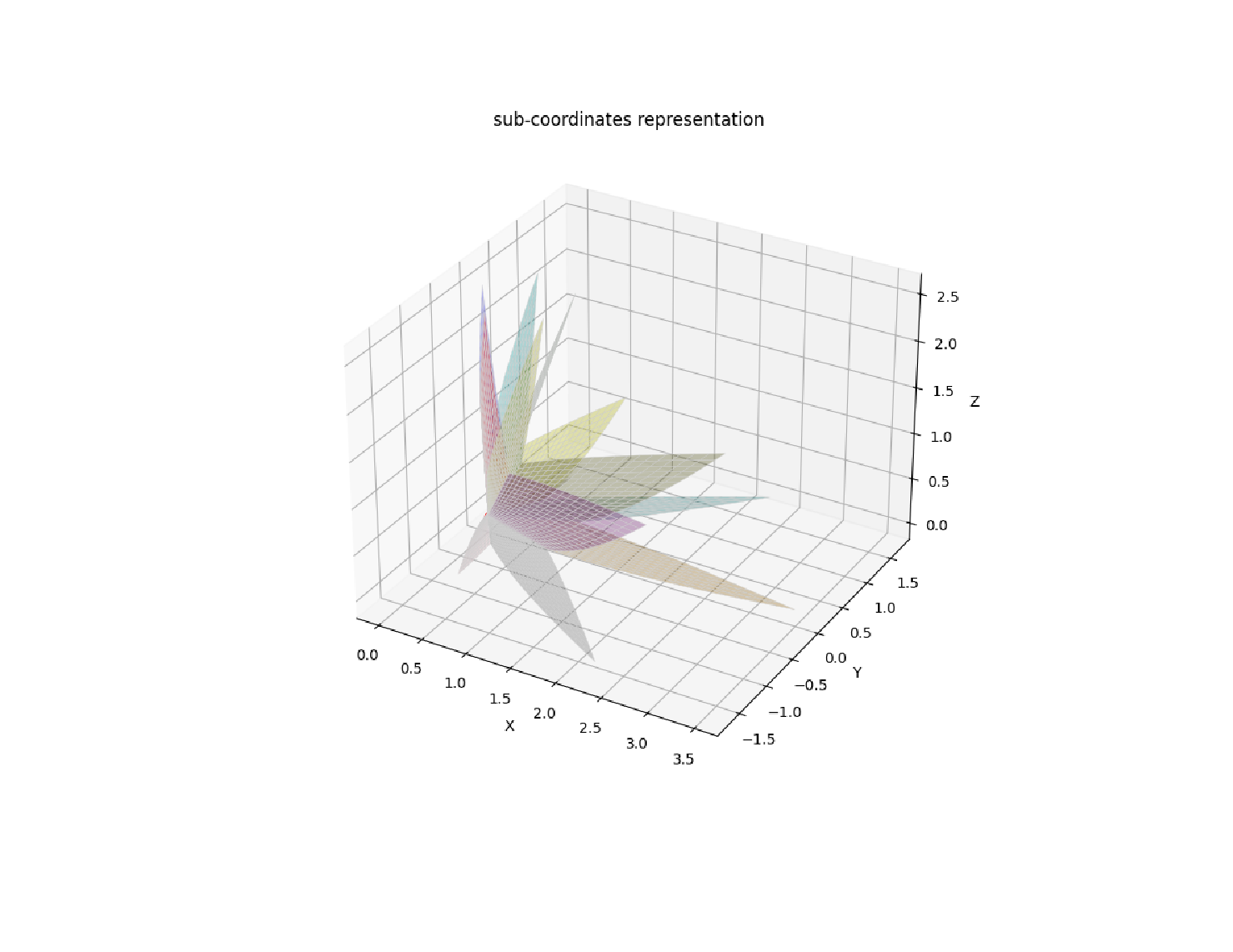}
\endminipage\hfill
\minipage{0.2\textwidth}%
    \includegraphics[width=\linewidth]{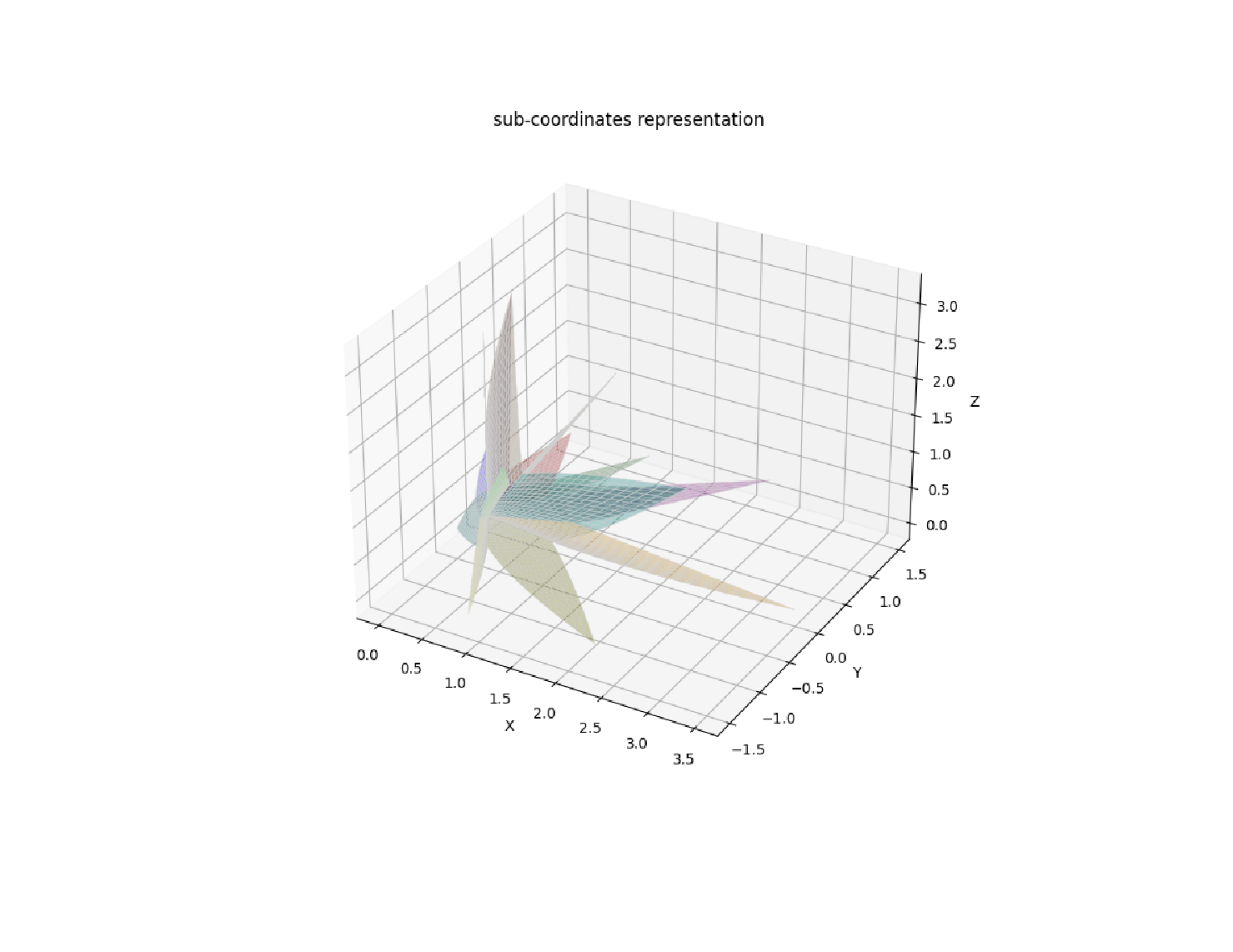}
\endminipage

\minipage{0.2\textwidth}
    \includegraphics[width=\linewidth]{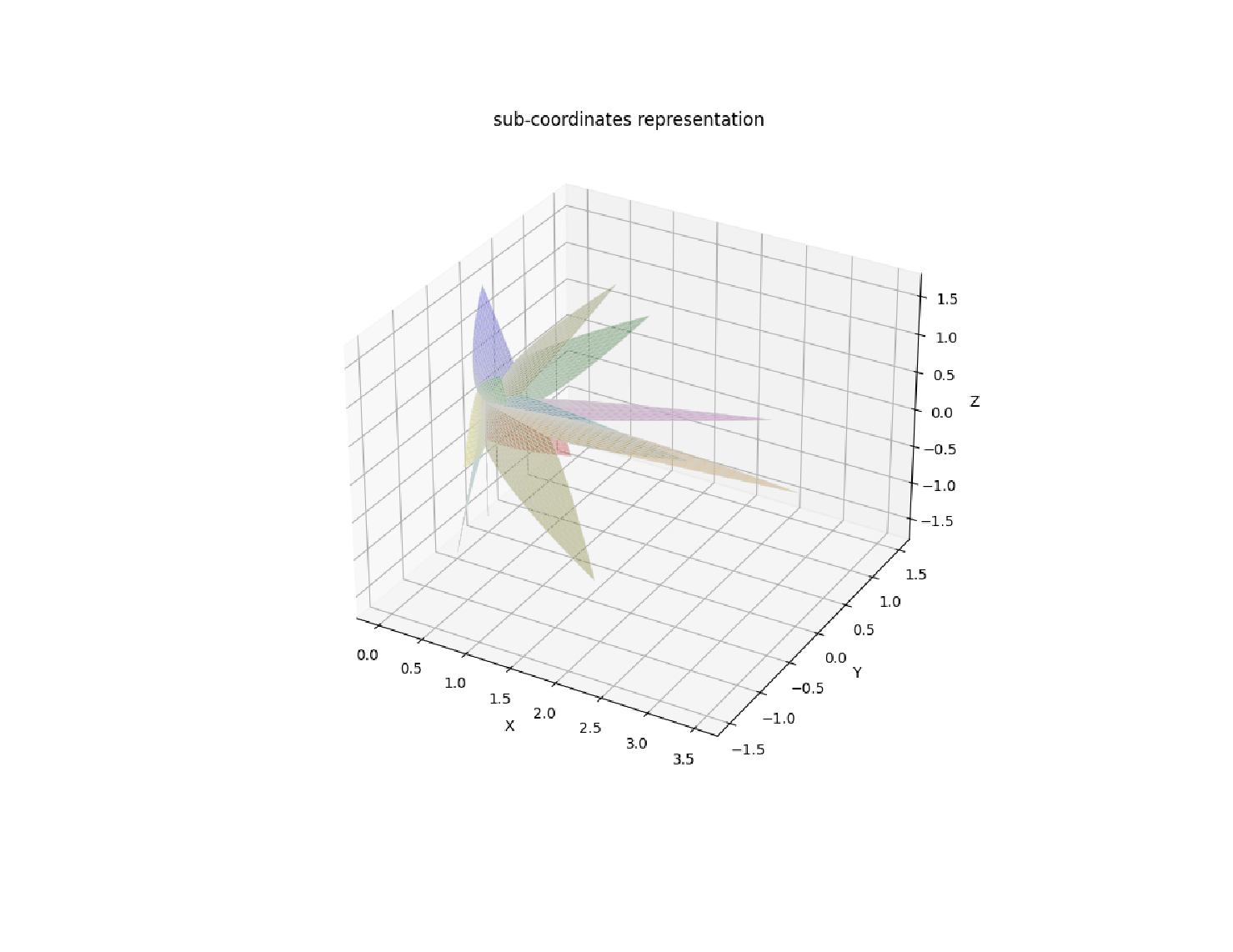}
\endminipage\hfill
\minipage{0.2\textwidth}
    \includegraphics[width=\linewidth]{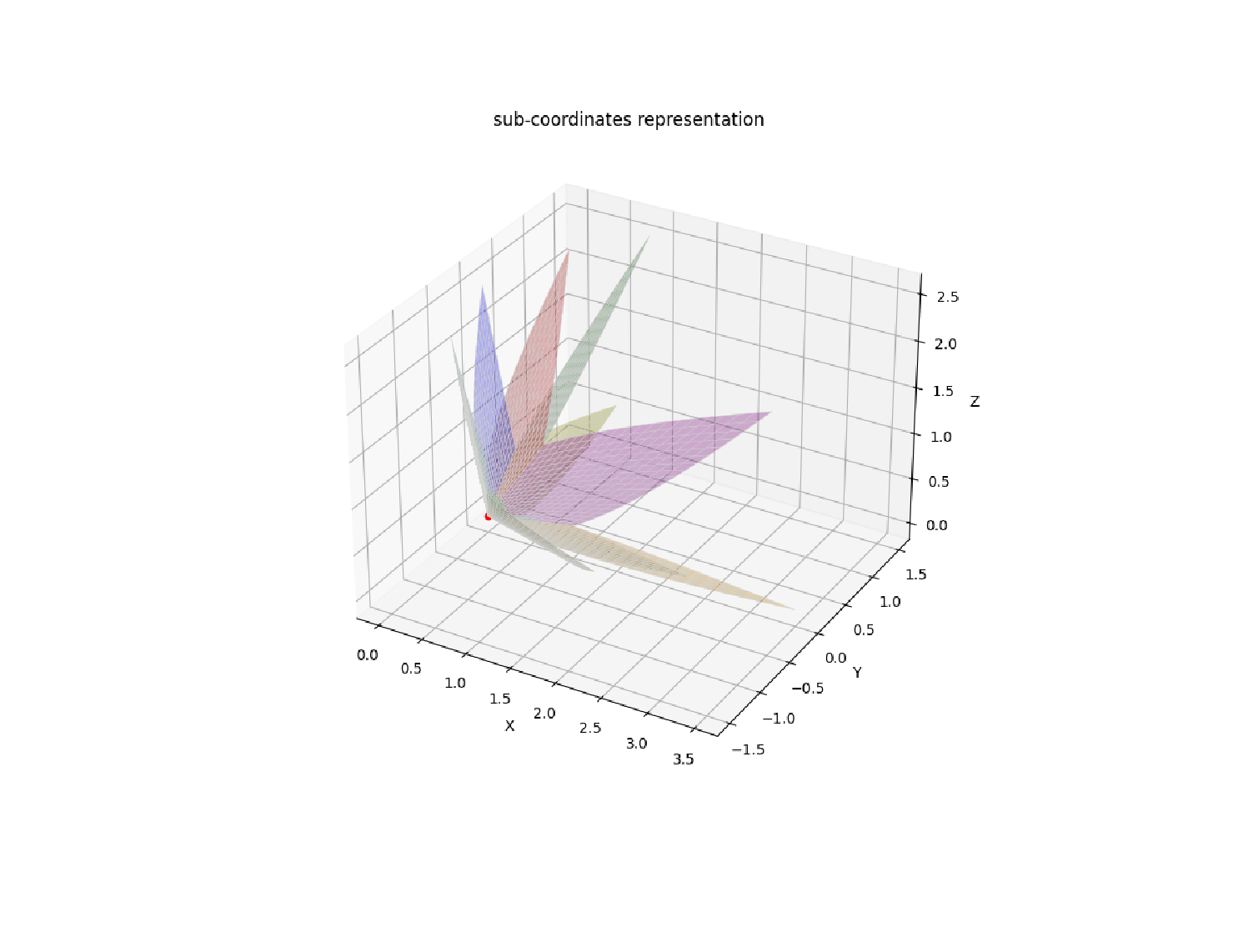}
\endminipage\hfill
\minipage{0.2\textwidth}%
    \includegraphics[width=\linewidth]{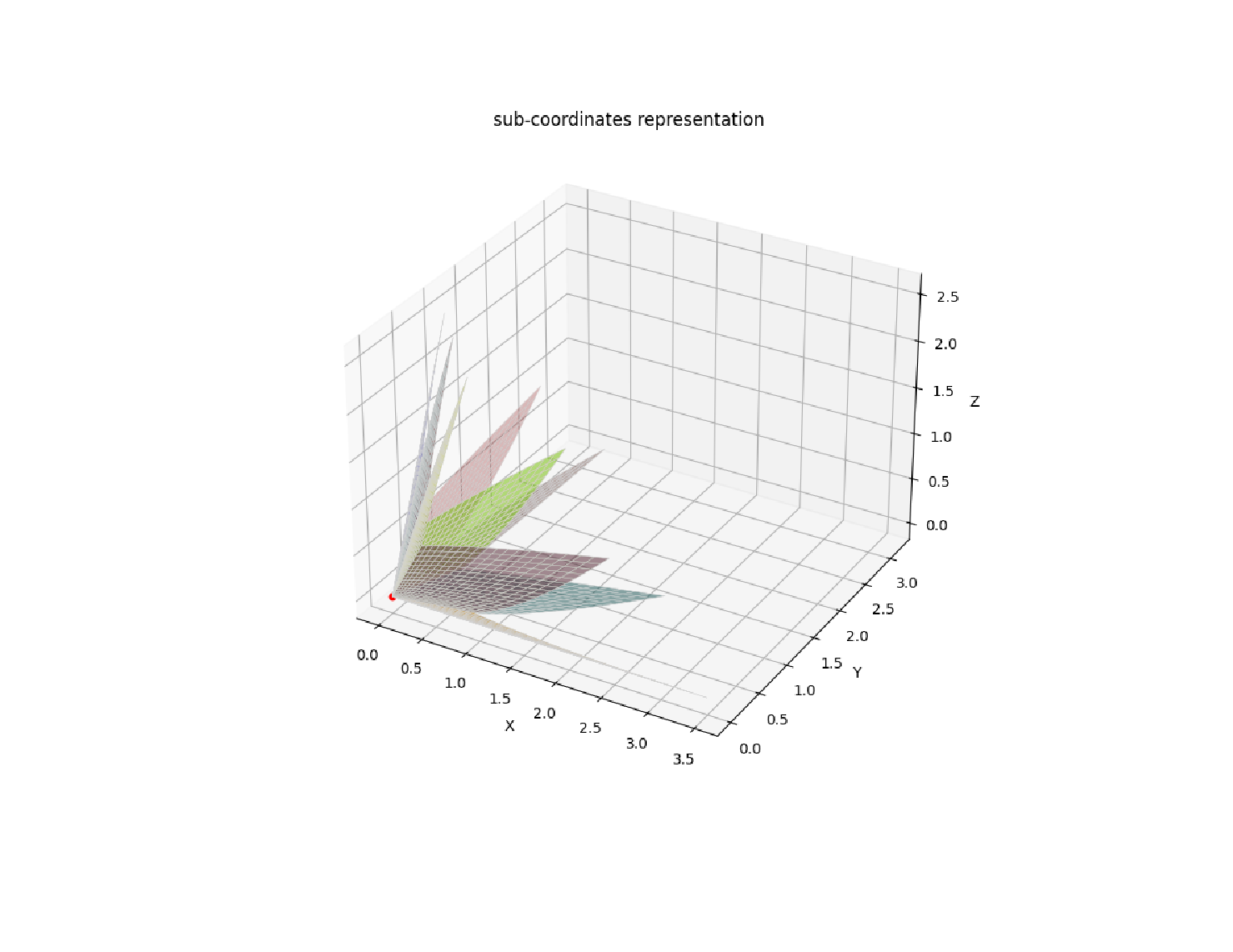}
\endminipage\hfill
\minipage{0.2\textwidth}
    \includegraphics[width=\linewidth]{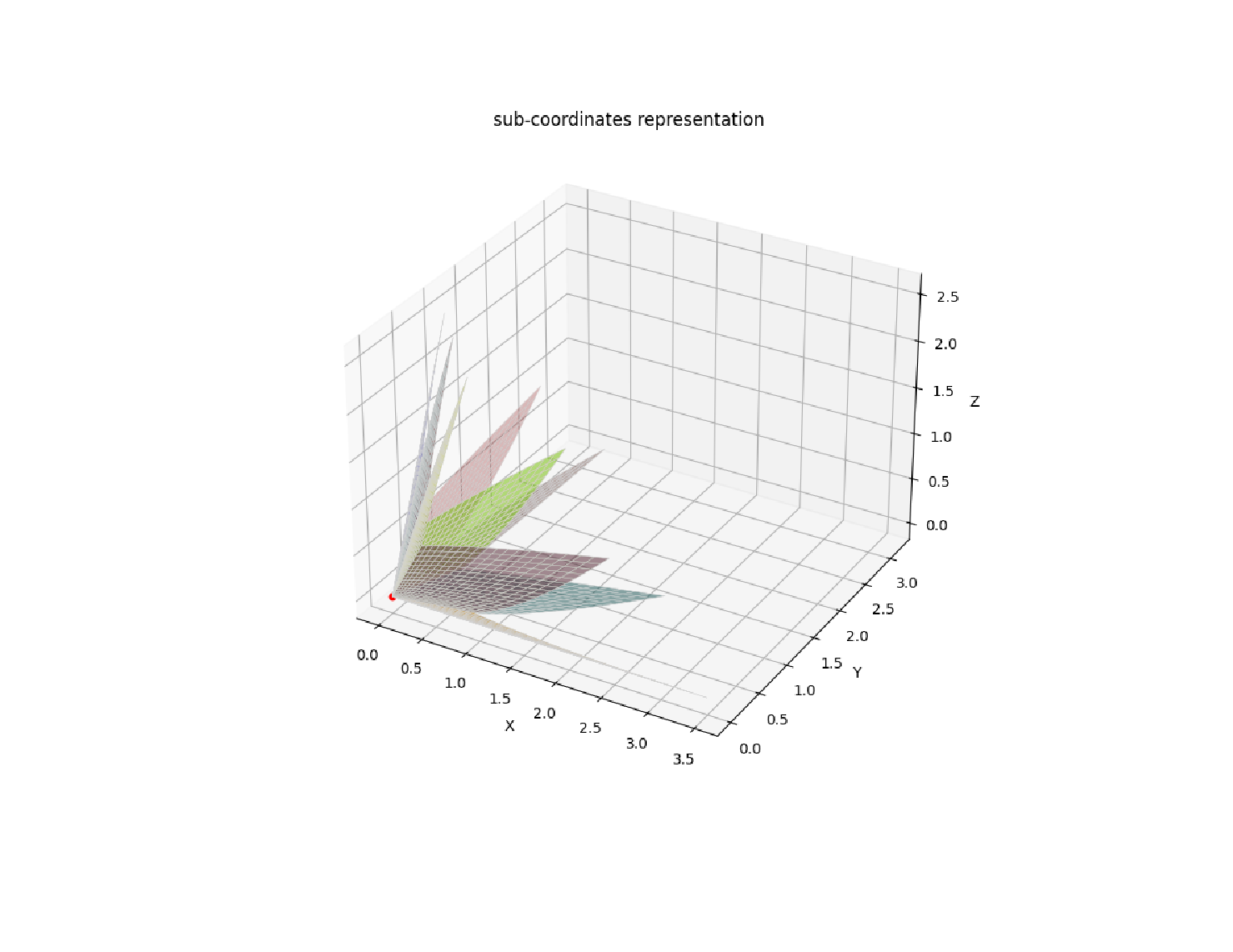}
\endminipage\hfill
\minipage{0.2\textwidth}%
    \includegraphics[width=\linewidth]{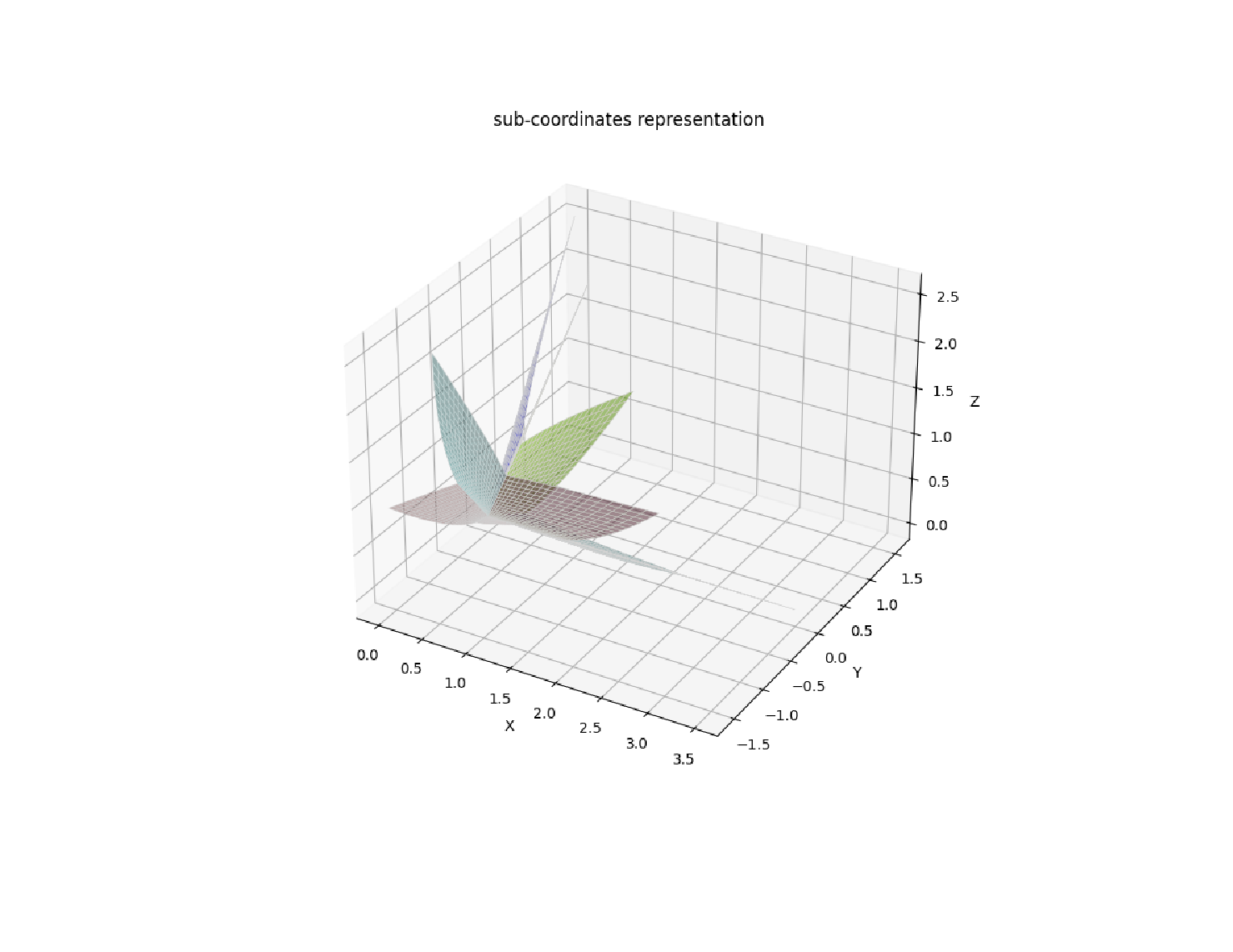}
\endminipage

\minipage{0.2\textwidth}
    \includegraphics[width=\linewidth]{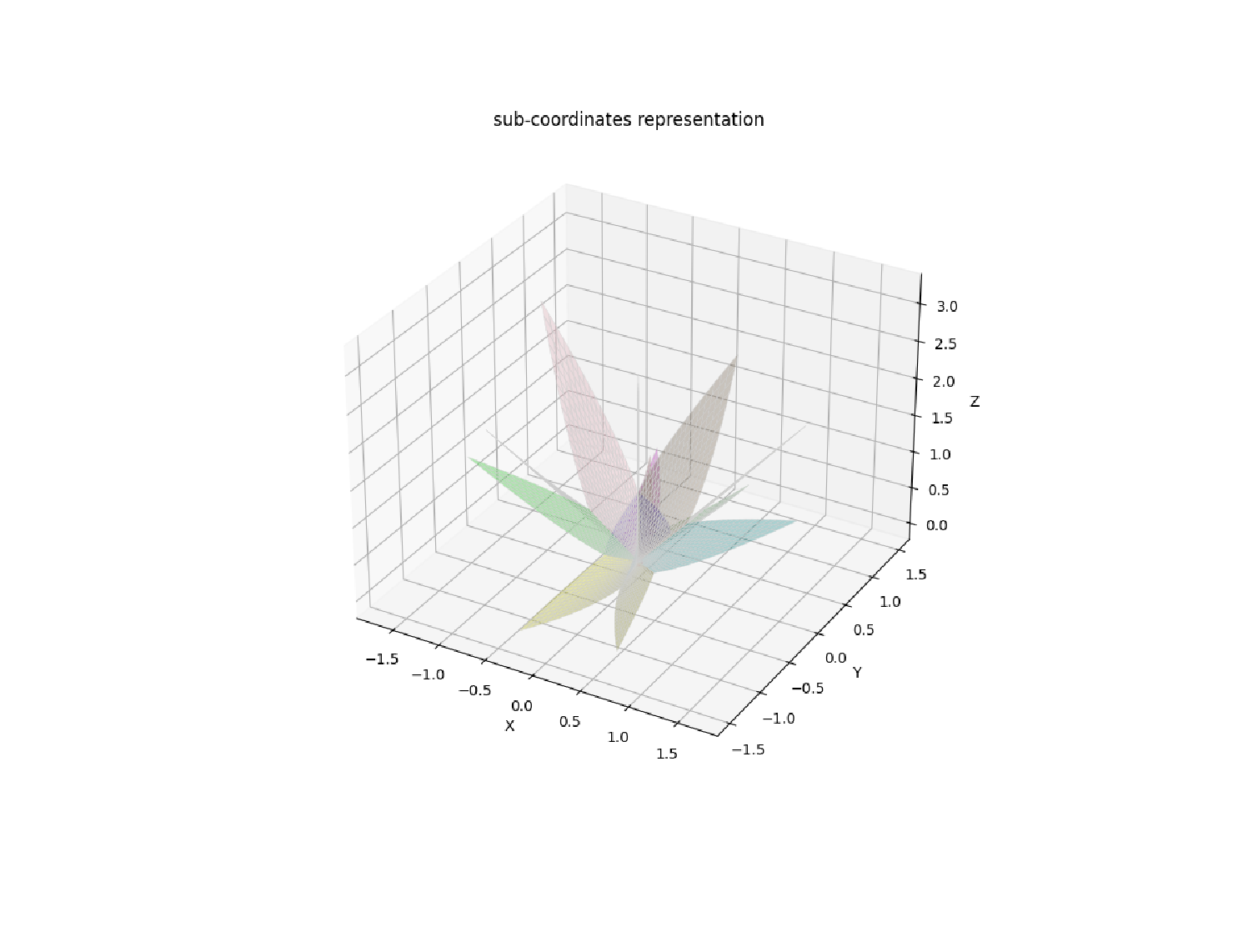}
\endminipage\hfill
\minipage{0.2\textwidth}
    \includegraphics[width=\linewidth]{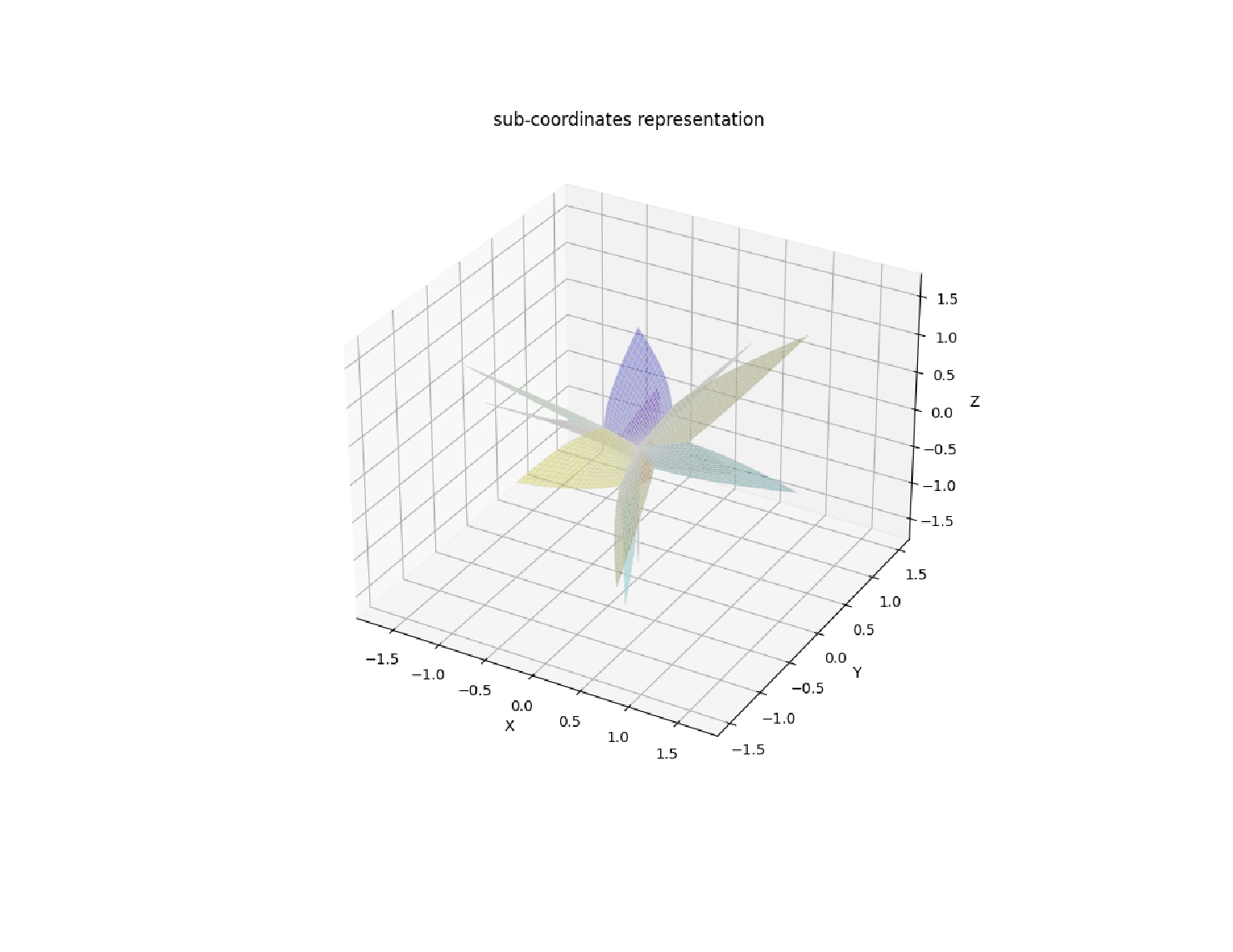}
\endminipage\hfill
\minipage{0.2\textwidth}%
    \includegraphics[width=\linewidth]{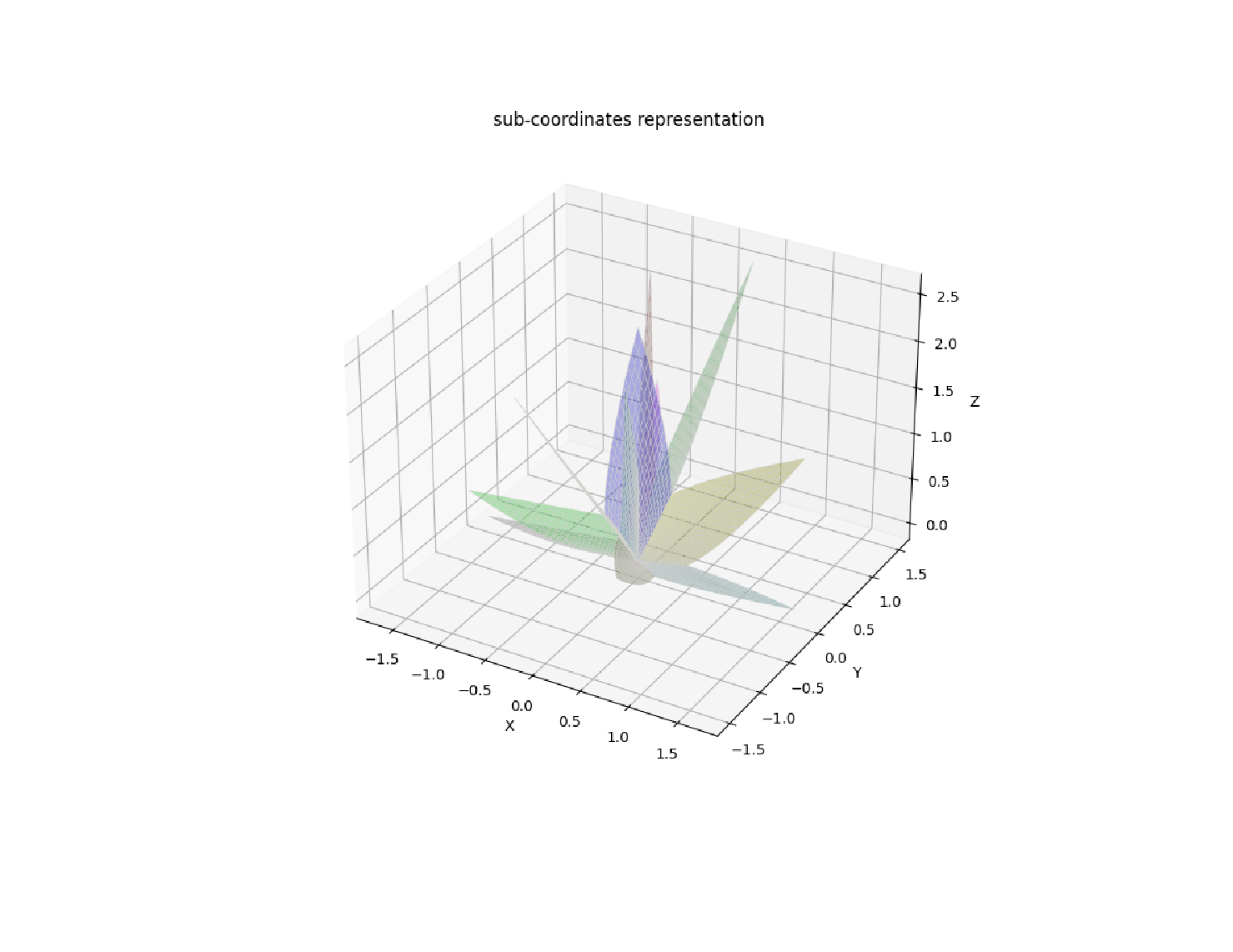}
\endminipage\hfill
\minipage{0.2\textwidth}
    \includegraphics[width=\linewidth]{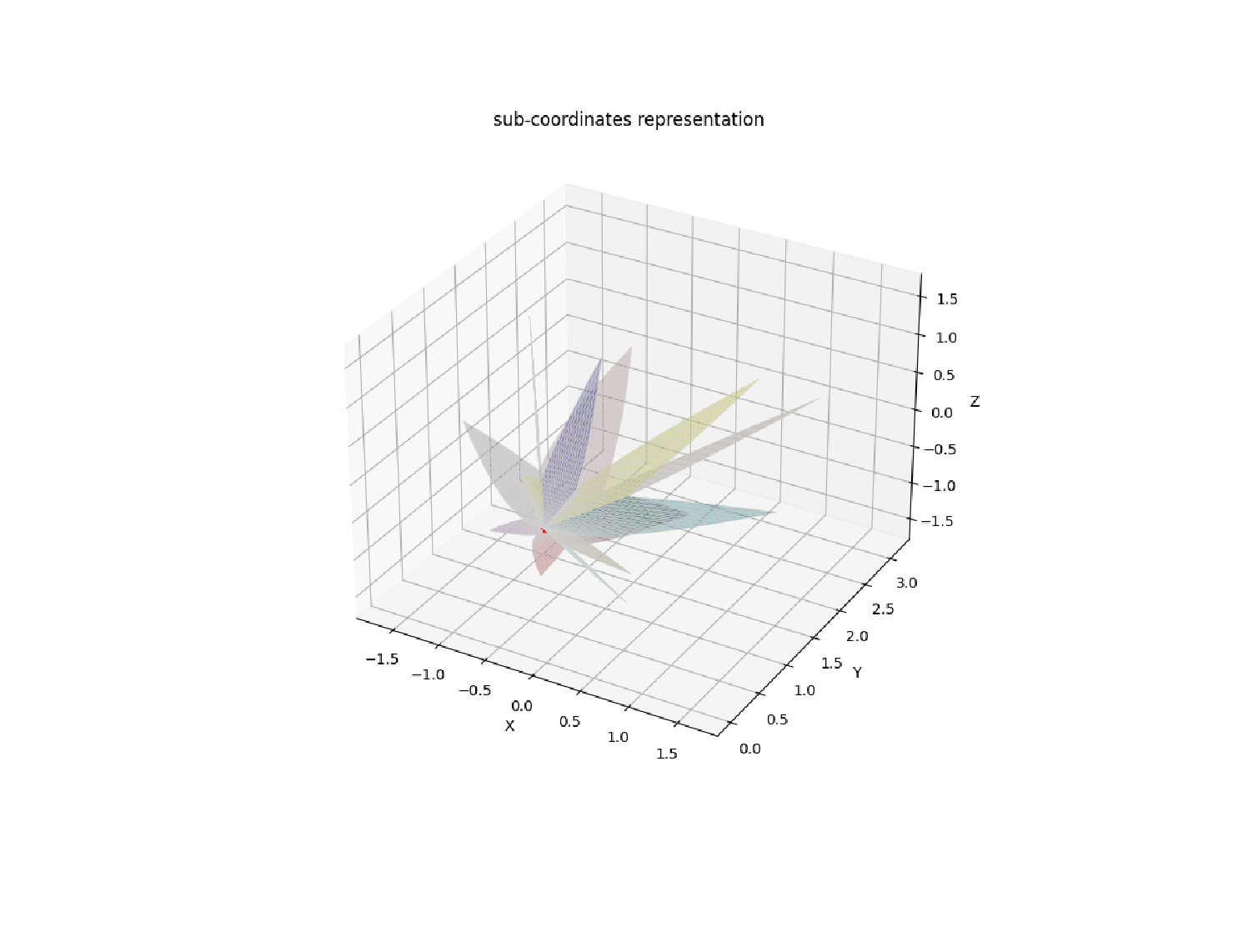}
\endminipage\hfill
\minipage{0.2\textwidth}%
    \includegraphics[width=\linewidth]{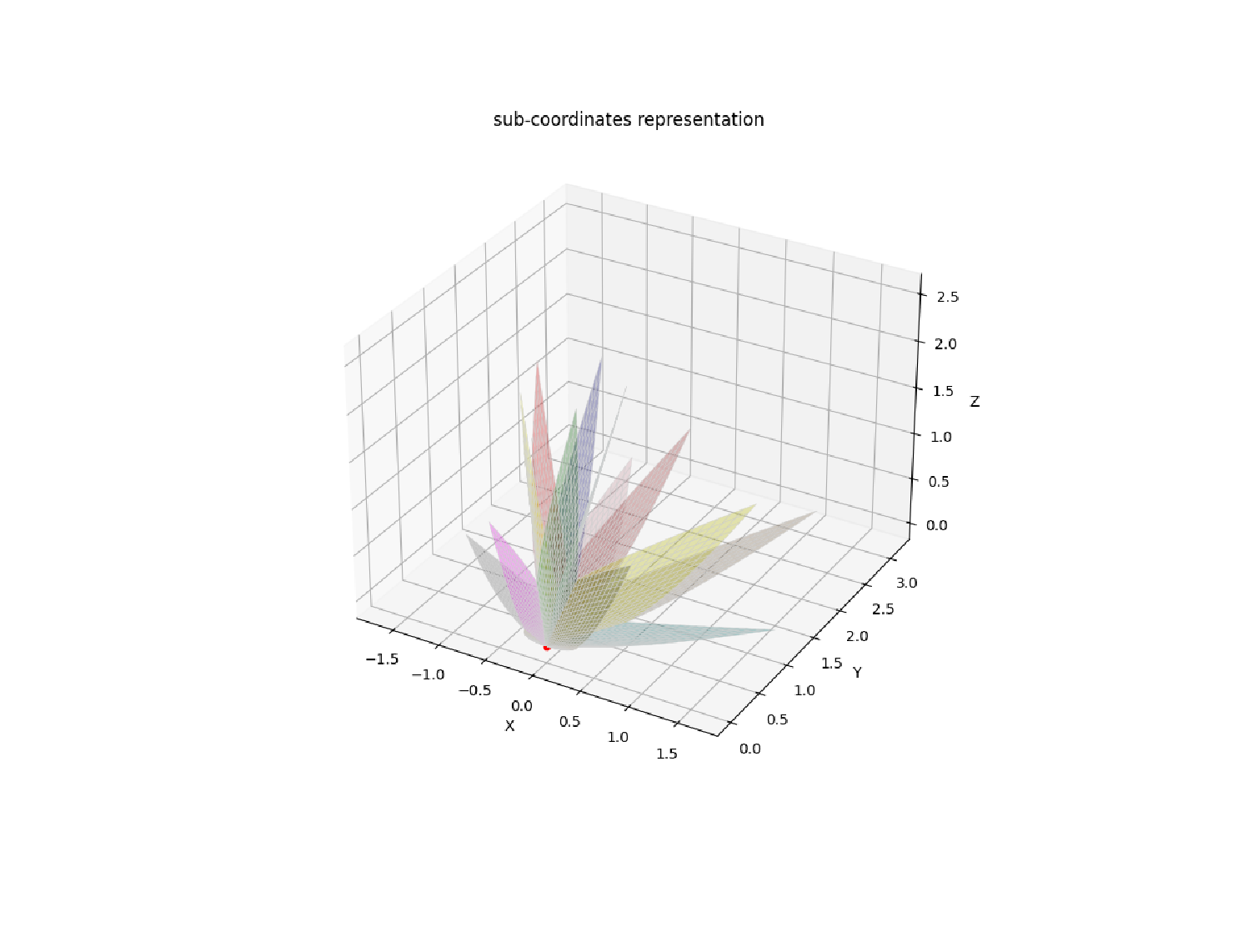}
\endminipage

\minipage{0.2\textwidth}
    \includegraphics[width=\linewidth]{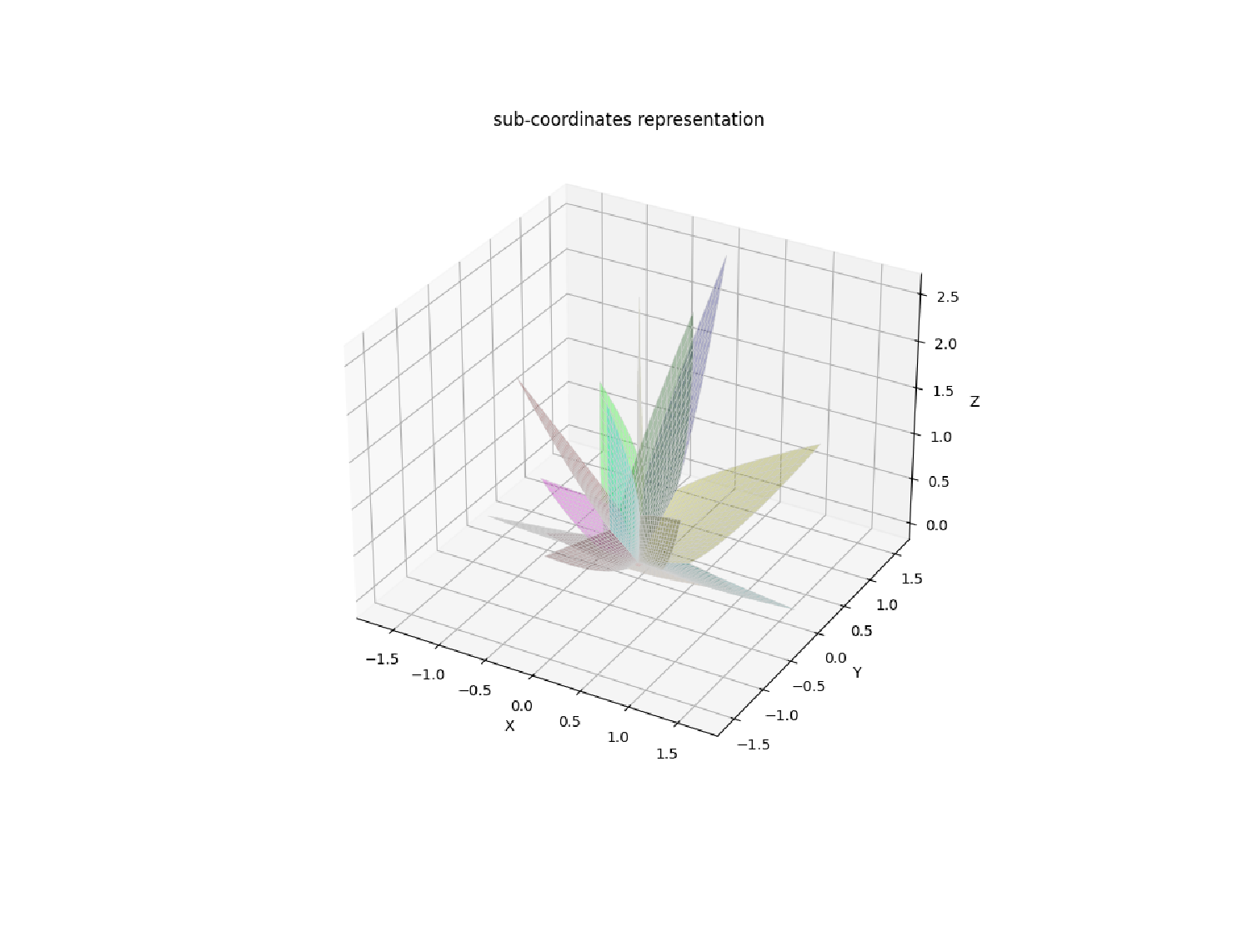}
\endminipage\hfill
\minipage{0.2\textwidth}
    \includegraphics[width=\linewidth]{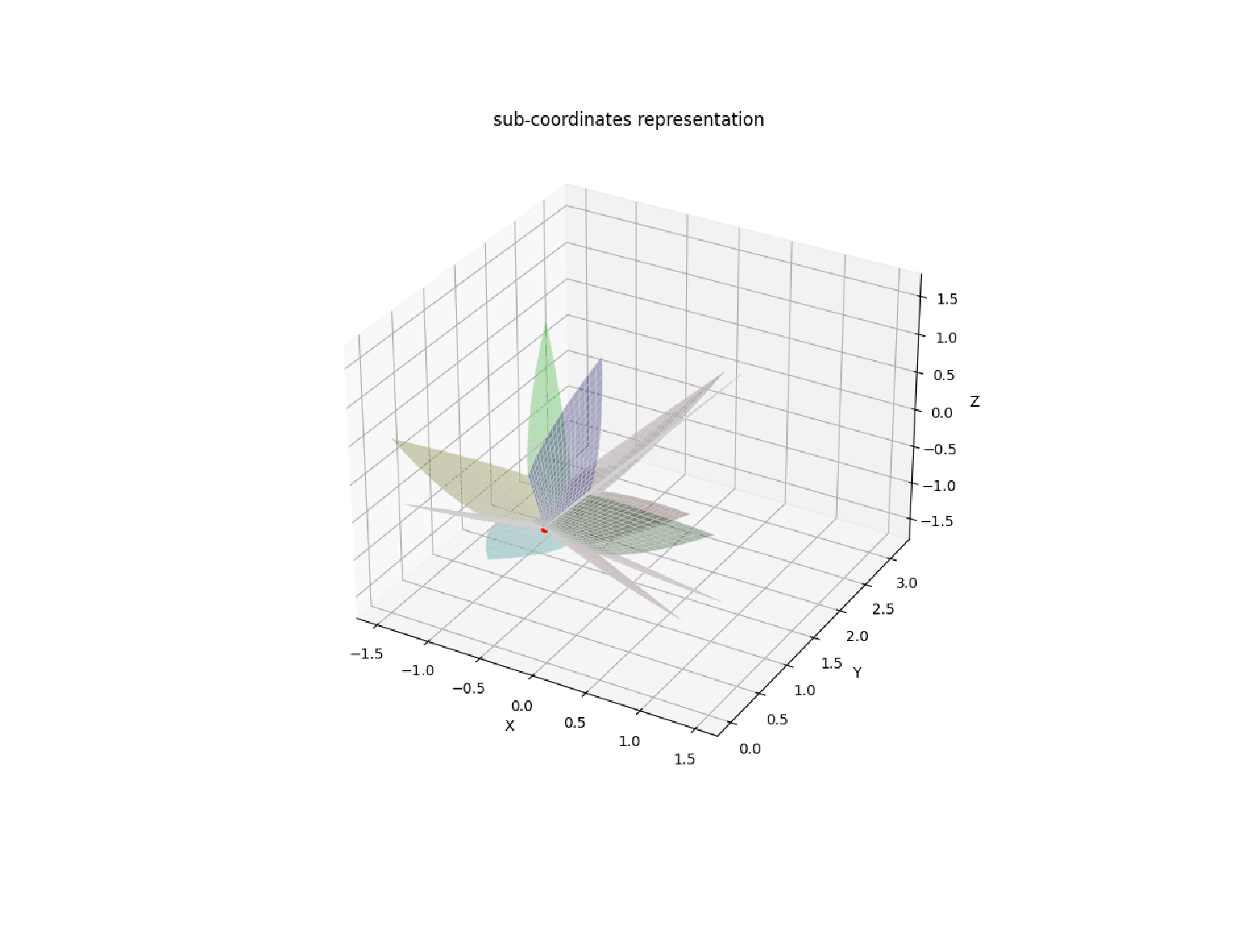}
\endminipage\hfill
\minipage{0.2\textwidth}%
    \includegraphics[width=\linewidth]{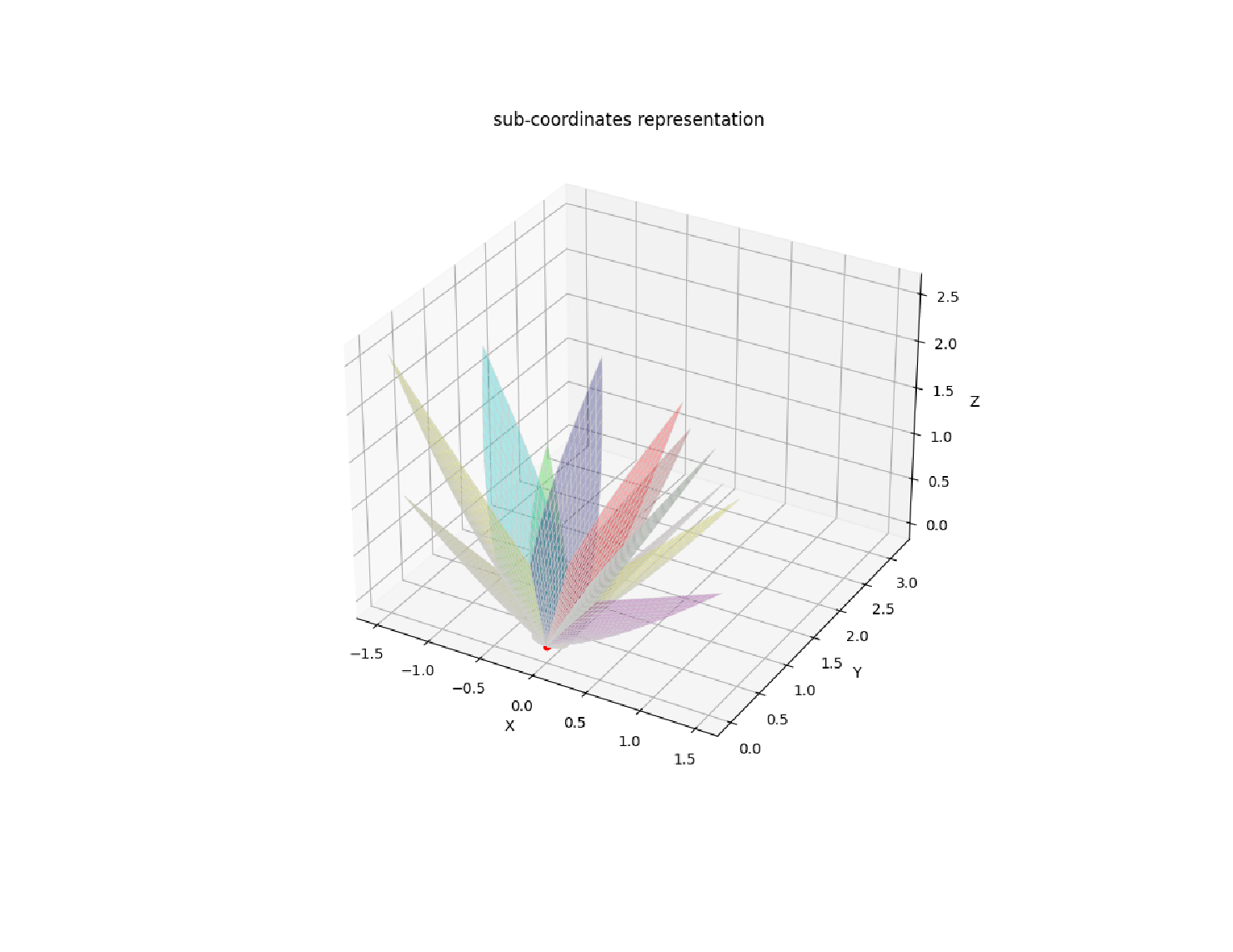}
\endminipage\hfill
\minipage{0.2\textwidth}
    \includegraphics[width=\linewidth]{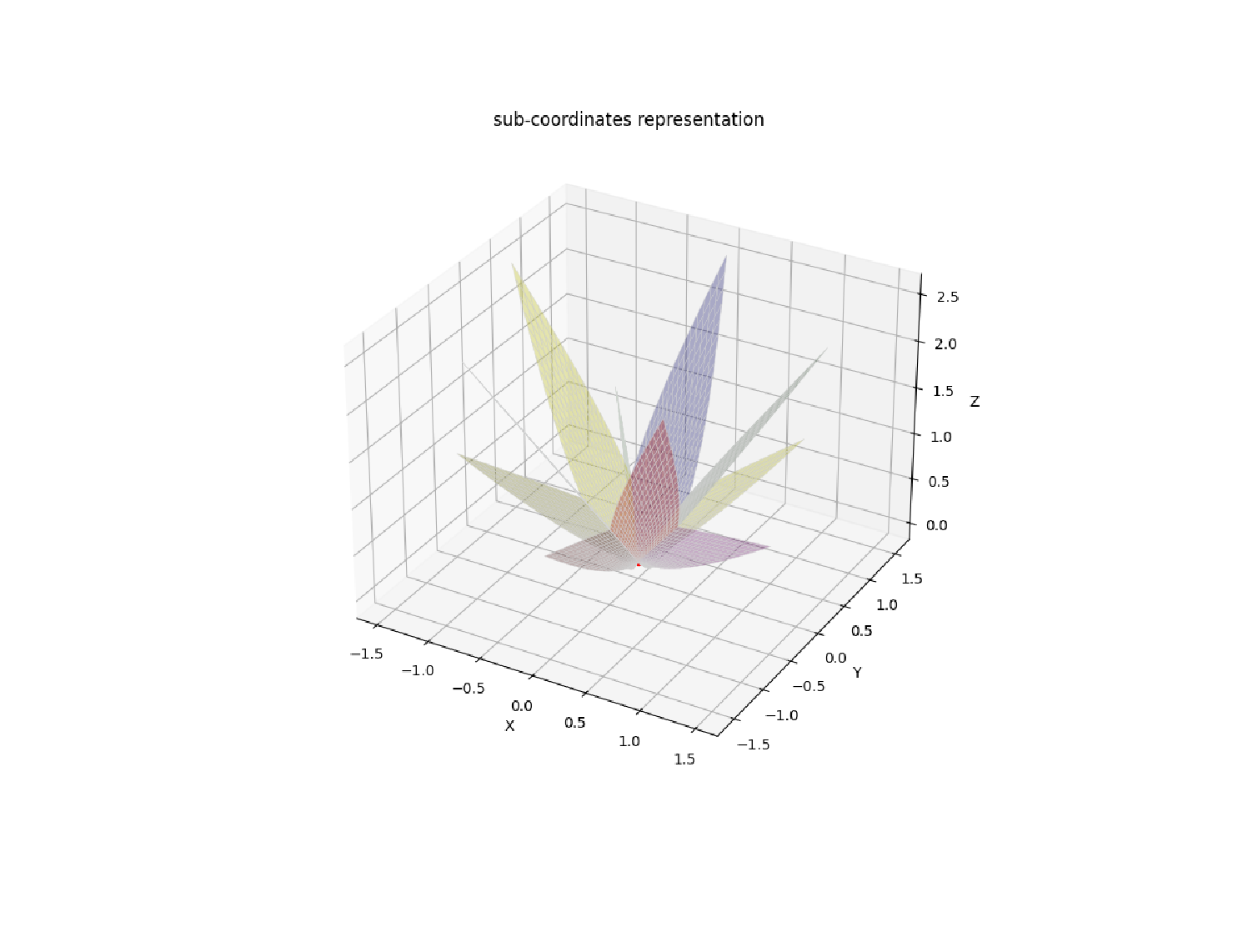}
\endminipage\hfill
\minipage{0.2\textwidth}%
    \includegraphics[width=\linewidth]{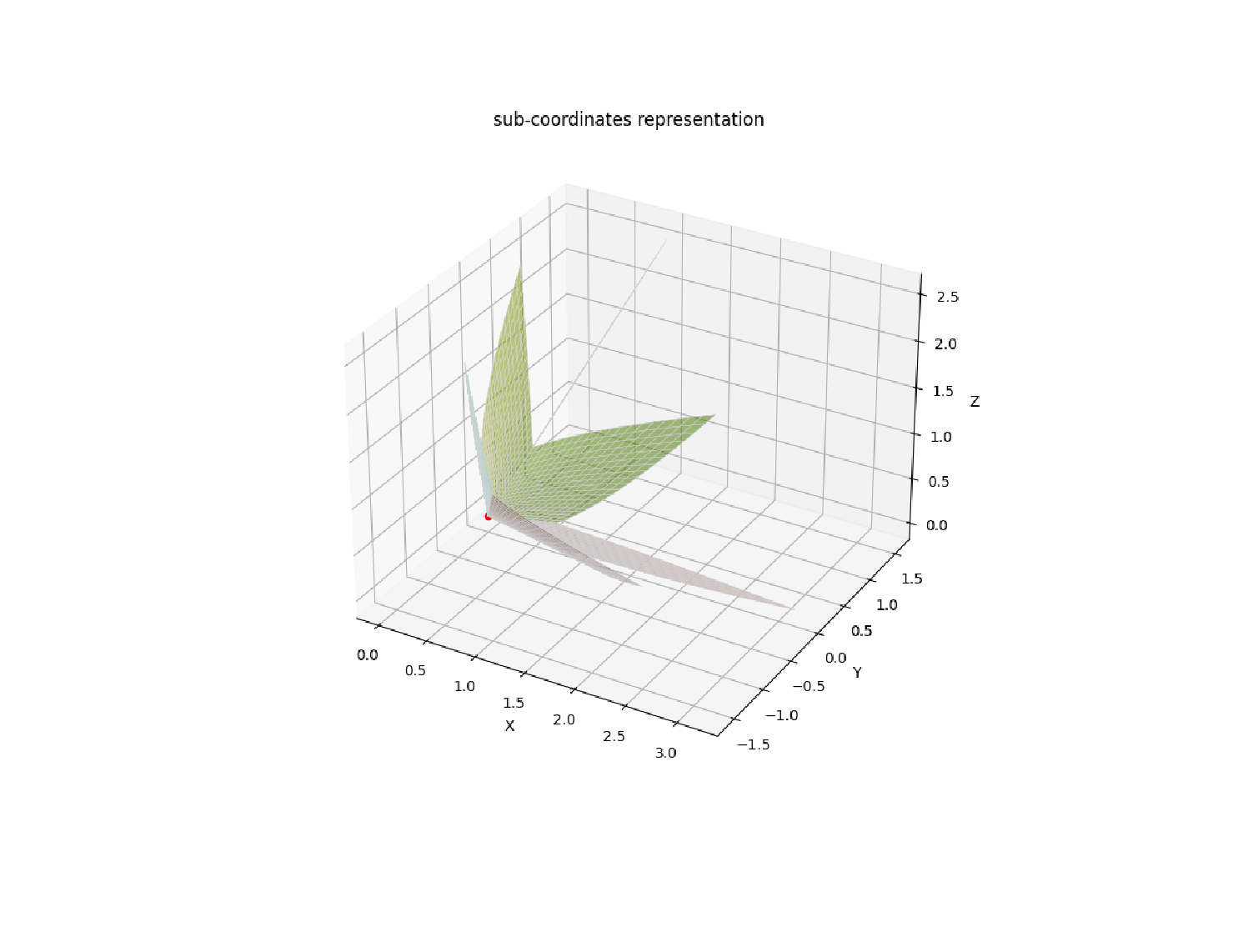}
\endminipage
\caption{The ``6 choose 3" 3D visualizations using the upper diagonal entries of the embedded Sped in the space of 3 by 3 symmetric matrices.}
\label{fig:20 views}
\end{figure}
\section{Acknowledgement} The authors acknowledge support from
award NSF-DMS:2311059, and are thankful to Robert L. Paige for his helpful suggestions.

\end{document}